\newcommand*{\ATLASLATEXPATH}{}
\pdfoutput=1 
\documentclass[UKenglish,texlive=2011,cernpreprint,txfonts,maketitle=true]{atlasdoc}

\usepackage[subfigure,subcaption]{\ATLASLATEXPATH atlaspackage}
\usepackage{\ATLASLATEXPATH atlasbiblatex}

\usepackage{\ATLASLATEXPATH atlascontribute}

\usepackage{\ATLASLATEXPATH atlasphysics}

\usepackage{url}
\usepackage{hyperref}
\usepackage{units}
\usepackage{textcomp}
\usepackage{multirow}

\graphicspath{{logos/}{figures/}}

\newcommand{\lumi}{20.3 fb$^{-1}$}
\newcommand{\pythia}{{\sc Pythia}}
\newcommand{\herwigpp}{{\sc Herwig++}}
\newcommand{\grav}{\ensuremath{G_{\mathrm{RS}}}}
\newcommand{\brit}{}

\newcommand{\locsig}{3.4}
\newcommand{\globsig}{2.5} 
\newcommand{\crosssection}{cross section}

\def\wpwz{\ensuremath{W' \rightarrow WZ}}
\def\mwp{\ensuremath{m_{W'}}}
\def\mjj{\ensuremath{m_{jj}}}
\def\kt{\ensuremath{k_t}}

\def\nexpvec{\ensuremath{\mathbf{n_{\mathrm exp}}}}
\def\nsigvec{\ensuremath{\mathbf{n_{\mathrm sig}}}}

\def\nbgvec{\ensuremath{\mathbf{n_{\mathrm bg}}}}
\def\nobsi{\ensuremath{n_{\mathrm obs}^i}}
\def\nexpi{\ensuremath{n_{\mathrm exp}^i}}

\def\nbgi{\ensuremath{n_{\mathrm bg}^i}}

\AtlasTitle{Search for high-mass diboson resonances with boson-tagged jets in proton--proton collisions at \boldmath $\sqrt{s}$ = 8\,TeV with the ATLAS detector}

\author{The ATLAS Collaboration}

\AtlasJournal{JHEP}
\PreprintIdNumber{CERN-PH-EP-2015-115}

\AtlasAbstract{ A search is performed for narrow resonances decaying into $WW$, $WZ$, or $ZZ$ boson pairs using 20.3 fb$^{-1}$ of proton--proton collision data at a centre-of-mass energy of $\sqrt{s}$ = 8 \TeV{} recorded with the ATLAS detector at the Large Hadron Collider. Diboson resonances with masses in the range from 1.3 to 3.0 TeV are sought after using the invariant mass distribution of dijets where both jets are tagged as a boson jet, compatible with a highly boosted $W$ or $Z$ boson decaying to quarks, using jet mass and substructure properties.
The  largest deviation from a smoothly falling background in the observed dijet invariant mass distribution occurs around 2 \tev{} in the $WZ$ channel, with a global significance of \globsig{} standard deviations.  Exclusion limits at the 95\% confidence level are set on the production \crosssection{} times branching ratio for the $WZ$ final state of  a new heavy gauge boson, $W'$, and for the $WW$ and $ZZ$ final states of Kaluza--Klein excitations of the graviton in a bulk Randall--Sundrum model, as a function of the resonance mass. $W'$ bosons with couplings predicted by the extended gauge model in the mass range from 1.3 to 1.5 \tev{} are excluded at 95\% confidence level. }

\hypersetup{pdftitle={ATLAS draft},pdfauthor={The ATLAS Collaboration}}

\begin{document}

\tableofcontents


\clearpage

\section{Introduction}
\label{sec:intro}

The substantial dataset of Large Hadron Collider (LHC) proton--proton ($pp$) collisions at $\sqrt{s} =$ 8~TeV collected by the ATLAS experiment provides a distinct opportunity to search for new heavy resonances at the \tev{} mass scale.  
This paper presents a search for narrow diboson resonances ($WW$, $WZ$ and $ZZ$) decaying to fully hadronic final states. The fully hadronic mode has a higher branching fraction than leptonic and semileptonic decay modes, and is therefore used to extend the reach of the search to the highest possible resonance masses.  

\Wboson{} and \Zboson{} bosons resulting from the decay of very massive resonances are highly boosted, so that each boson's hadronic decay products are reconstructed as a single jet. The signature of the heavy resonance decay is thus a resonance structure in the dijet invariant mass spectrum. The dominant background for this search is due to dijet events from QCD processes, which produce a smoothly falling spectrum without resonance structures.  To cope with this large background, jets are selected using a boson tagging procedure based on a reclustering-mass-drop filter (BDRS-A, similar to the method introduced in ref.~\cite{bdrs}), jet mass, and further substructure properties.  The tagging procedure strongly suppresses the dijet background, although these QCD processes still overwhelm the expected backgrounds from single boson production with one or more jets, Standard Model (SM) diboson production, single-top and top-pair production.  As all of these background sources produce dijet invariant mass distributions without resonance peaks, the expected background in the search is modelled by a fit to a smoothly falling distribution.

Diboson resonances are predicted in several extensions to the SM, such as technicolour ~\cite{Eichten:2007sx,Catterall:2011zf,Andersen:2011yj}, warped extra dimensions ~\cite{Randall:1999ee,Randall:1999vf,Davoudiasl:2000wi}, and Grand Unified Theories ~\cite{Pati:1974yy,Georgi:1974sy,Georgi:1974my,Fritzsch:1974nn}.  To assess the sensitivity of the search, to \brit{optimise} the event selection, and for comparison with data, two specific benchmark models are used:
an extended gauge model (EGM)  $W' \to WZ$ where the spin-1 $W'$ gauge boson has a modified coupling to the SM \Wboson{} and \Zboson{} bosons~\cite{Altarelli1989,EHLQ1984ref,Quigg2013},
and a spin-2 graviton, $G_{RS} \to WW$ or $ZZ$, a Kaluza--Klein mode~\cite{Randall:1999ee,KKmodesHLZ} of the bulk Randall--Sundrum (RS) graviton~\cite{ADPS2007,AAS2008,AS2008}. 

The CMS collaboration has performed a search for diboson resonances with the fully hadronic final state~\cite{CMSjetref} of comparable sensitivity to the one presented in this article. In this search, the EGM $W' \to WZ$ with masses below 1.7 \TeV{} and $\grav$ of the original RS model decaying to $WW$ with masses below 1.2 TeV  are excluded at 95\% confidence level (CL). The CMS collaboration has also published upper limits on the production of generic diboson resonances using semileptonic final states~\cite{Khachatryan:2014gha}.  Using the  $\ell\ell q\bar{q}$ final state, the ATLAS collaboration has excluded at 95\% CL a bulk $\grav \to ZZ$ with mass below 740 \gev{}~\cite{atlas:llqq_2012}.  For narrow resonances decaying exclusively to $WZ$ or $WW$, 
the sensitivity of the ATLAS search in the $\ell \nu q\bar{q}$ channel~\cite{atlas:lnuqq_2012} is comparable to that of the search presented here in the mass range from 1.3 to 2.5 \tev{}. That search has also excluded a bulk $\grav \to WW$ with mass below 760 \gev{}.

\section{ATLAS detector and data sample}
The ATLAS detector \cite{atlas:detector} surrounds nearly the entire solid angle around the ATLAS collision point.  It has an approximately cylindrical geometry and consists of an inner tracking detector surrounded by electromagnetic and hadronic calorimeters and a muon spectrometer.  The tracking detector is placed within a 2~T axial magnetic field provided by a superconducting solenoid and measures charged-particle trajectories with pixel and silicon microstrip detectors that cover the pseudorapidity\footnote{ATLAS uses a right-handed coordinate system with its origin at the nominal interaction point (IP) in the centre of the detector and the $z$-axis along the beam pipe. The $x$-axis points from the IP to the centre of the LHC ring, and the $y$-axis points upward. Cylindrical coordinates $(r,\phi)$ are used in the transverse plane, $\phi$ being the azimuthal angle around the beam pipe. The pseudorapidity is defined in terms of the polar angle $\theta$ as $\eta=-\ln\tan(\theta/2).$} range $|\eta| < 2.5$, and with a straw-tube transition radiation tracker covering $|\eta| < 2.0$.

A high-granularity electromagnetic and hadronic calorimeter system measures the jets in this analysis.  The electromagnetic calorimeter is a liquid-argon (LAr) sampling calorimeter with lead absorbers, spanning $|\eta| < 3.2$ with barrel and end-cap sections.  The three-layer central hadronic calorimeter comprises scintillator tiles with steel absorbers and extends to $|\eta| = 1.7$.  The hadronic end-cap calorimeters measure particles in the region $1.5 < |\eta| < 3.2$ using liquid argon with copper absorber.  The forward calorimeters cover $3.1 < |\eta| < 4.9$, using LAr/copper modules for electromagnetic energy measurements and LAr/tungsten modules to measure hadronic energy.

Events are recorded in ATLAS if they satisfy a three-level trigger requirement. The level-1 trigger detects jet and particle signatures in the calorimeter and muon systems with a fixed latency of 2.5 $\mu$s, and is designed to reduce the event rate to less than 75 kHz. Jets are identified at level-1 with a sliding-window algorithm, searching for local maxima in square regions with size $\Delta \eta \times \Delta \phi = 0.8 \times 0.8$.  The subsequent high-level trigger consists of two stages of software-based trigger filters which reduce the event rate to a few hundred Hz.
Events used in this search satisfy a single-jet trigger requirement, based on at least one jet reconstructed at each trigger level.  At the first filtering stage of the high-level trigger, jet candidates are reconstructed from calorimeter cells using a cone algorithm with small radius, $R$=0.4.  The final filter in the high-level trigger requires a jet to satisfy a higher transverse momentum (\pt{}) threshold, reconstructed with the anti-\kt{} algorithm \cite{antikt} and a large radius parameter ($R$=1.0). 

This search is performed using the dataset collected in 2012 from 8 \TeV{} LHC $pp$ collisions using a single-jet trigger with a nominal \pt{} threshold of 360 \GeV{}. The integrated luminosity of this dataset after requiring good beam and detector conditions is \lumi, with a relative uncertainty of $\pm 2.8\%$. The uncertainty is derived following the methodology detailed in ref.~\cite{atlas:lumi2011}. 

\section{Simulated data samples}\label{sec:sim}

The leading-order Monte Carlo (MC) generator \pythia{} 8.170~\cite{pythia8}  is used to model \wpwz\ events in order to determine and optimise the sensitivity of this search.
\pythia{} 8 uses the \pt-ordered showering introduced in \pythia{} 6.3~\cite{Sjostrand:2004ef,pythia}, and interleaves multiple parton interactions with both initial- and final-state radiation.  The samples generated for this analysis use MSTW2008~\cite{mstw} parton distribution functions (PDFs), with parton shower parameters tuned to ATLAS underlying-event data \cite{atlas:a2tune}.  Hadronisation is based on the Lund string fragmentation framework~\cite{lund}.  An additional set of $W'$ samples generated with \pythia{} for the hard scattering interaction and \herwigpp{}~\cite{Bahr:2008pv} for parton showering and hadronisation is used to assess systematic uncertainties on the signal efficiency due to uncertainties on the parton shower and hadronisation model.  These samples use angular-ordered showering and cluster hadronisation.

The $W'$ boson samples are generated for different resonance masses, covering the range $1.3 \leq m_{W'} \leq 3.0$ \TeV{} in 100 \GeV{} intervals. The $W'$ is required to decay to a \Wboson{} and a \Zboson{} boson, which are both forced to decay hadronically.  The \crosssection{} times branching ratio as well as the resonance width for the samples listed in table~\ref{tab:simwprime} are calculated by \pythia{} 8 assuming EGM couplings~\cite{Altarelli1989} for the $W'$. In particular, the $W'$ coupling to $WZ$ is equal to that of the $W$ coupling scaled by $c \times \left({m_W}/{m_{W'}}\right)^2$, where $c$ is a coupling scaling factor of order one which is set to unity for the samples generated here.  The partial width of \wpwz\ decays thus scales linearly with \mwp, leading to a narrow width over the entire accessible mass range.  Because of the anti-quark parton distribution functions involved in the production,  a significant part of the $W'$  \crosssection{} for large $W'$ masses is due to off-shell interactions which produce a low-mass tail in the $W'$ mass spectrum. The relative size of the low-mass tail increases with the $W'$ mass: the fraction of events with a diboson mass below 20\% of the pole mass of the $W'$ increases from  10\% for $m_{W'} = 1.3 \tev{}$ to 22\% for $m_{W'} = 2.0 \tev{}$ and to 65\% for $m_{W'} = 3.0 \tev{}$.

An extended RS model with a warped extra dimension is used for the excited graviton benchmarks. In this model the SM fields are allowed to propagate in the warped extra dimension \cite{ADPS2007}, avoiding the constraints on the original RS model from limits on flavour-changing neutral currents and electroweak precision measurements. The model is characterised by a dimensionless coupling constant $k/\bar{M}_\mathrm{Pl} \sim 1$, where $k$ is the curvature of the warped extra dimension and $\bar{M}_\mathrm{Pl}$ is the reduced Planck mass. The RS excited graviton samples are generated with CalcHEP 3.4~\cite{calchep} setting $k/\bar{M}_\mathrm{Pl} = 1$, covering the resonance mass range $1.3 \leq m_{\grav} \leq 3.0$ \TeV{} in 100 \GeV{} intervals.  The graviton resonance is decayed to $WW$ or $ZZ$, and the resulting \Wboson{} or \Zboson{} bosons are forced to decay hadronically. The \crosssection{} times branching ratio as well as the resonance width calculated by CalcHEP for the RS model are listed in table~\ref{tab:simwprime}.  Events are generated using CTEQ6L1~\cite{cteq6l-1} PDFs, and use \pythia{} 8 for the parton shower and \brit{hadronisation}. 

To characterise the expected dijet invariant mass spectrum in the mass range 1.3--3.0 \tev{}, simulated QCD dijet events, diboson events, and single $W$ or $Z$ bosons produced with jets are used. Contributions from SM diboson events are expected to account for approximately 6\% of the selected sample, and single boson production is expected to contribute less than 2\%. Contributions from $t\bar{t}$ production, studied using {MC@NLO}~\cite{mcatnlo} and HERWIG~\cite{herwig} showering, were found to be negligible and are not considered further.

QCD dijet events are produced with \pythia{} 8 and the CT10~\cite{ct10} PDFs and the $\Wboson{}/\Zboson{}+\text{jets}$ samples are produced with \pythia{} 8 and CTEQ6L1 PDFs.  Diboson events are produced at the generator level with {POWHEG}~\cite{powheg}, using \pythia{} for the soft parton shower. 
The samples of single \Wboson{} or \Zboson{} bosons produced with jets are further used to determine a scale factor for the efficiency of the boson tagging selection, by comparing the boson tagging efficiencies between simulation and collision data in a $W/Z$+jets-enriched sample.  

The final-state particles produced by the generators are propagated through a detailed detector simulation~\cite{atlas:sim} based on GEANT4~\cite{geant}.  The average number of $pp$ interactions per bunch crossing  was approximately 20 while the collision data were collected.  The expected contribution from these additional minimum-bias $pp$ interactions is accounted for by overlaying additional minimum-bias events generated with \pythia{} 8, matching the distribution of the number of interactions per bunch crossing observed in collision data.   Simulated events are then reconstructed with the same algorithms run on collision data.

\begin{table}[!pht]
\caption{The resonance width ($\Gamma$) and the product of \crosssection{}s and branching ratios (BR) to four-quark final states used in modelling $\wpwz$, $\grav\rightarrow WW$, and $\grav \rightarrow ZZ$, for several values of resonance pole masses ($m$).  The fraction of events in which the invariant mass of the $W'$ or $\grav{}$ decay products lies within 10\% of the nominal resonance mass ($f_\text{10\%}$) is also displayed.}\label{tab:simwprime}
\begin{center}
\begin{tabular}{ccc|rr|rr|rr}
\hline\hline
                           &                           &                                  &\multicolumn{2}{c|}{$\wpwz$} & \multicolumn{2}{c|}{$\grav\rightarrow WW$} & \multicolumn{2}{c}{$\grav\rightarrow ZZ$} \\

 $m$                   & $\Gamma_{W'}$ &    $\Gamma_{\grav}$ & $\sigma\times\!\mathrm{BR}$ & $f_\text{10\%}$  &                                                                                          $\sigma\times\!\mathrm{BR}$ & $f_\text{10\%}$ &   $\sigma\times\!\mathrm{BR}$ & $f_\text{10\%}$ \\

$[\text{TeV}]$     & $[\text{GeV}]$  &   $ [\text{GeV}]$ & [fb]\phantom{00}  & & [fb]\phantom{00} & &  [fb]\phantom{00} &\\
\hline
 1.3                     &  47 & 76 & 19.1\phantom{00} & 0.83 & 0.73\phantom{000} & 0.85 & 0.37\phantom{000} & 0.84 \\
 1.6                     & 58  & 96 & 6.04\phantom{0} & 0.79 &  0.14\phantom{000} & 0.83 & 0.071\phantom{00} & 0.84 \\
 2.0                     &  72  & 123  &     1.50\phantom{0}   & 0.72 & 0.022\phantom{00}  & 0.83	 & 0.010\phantom{00} & 0.82 \\
 2.5                     & 91 & 155  & 0.31\phantom{0} &  0.54 & 0.0025\phantom{0} & 0.78 & 0.0011\phantom{0} & 0.78 \\
 3.0                     & 109    &187          &     0.088\phantom{}  &  0.31 &	0.00034	  & 0.72 & 0.00017 &  0.71 \\ 
\hline\hline
\end{tabular}
\end{center}
\end{table}

\section{Boson jet identification}
\label{sec:bosid}

\newcommand{\filt}{\mathrm{f}}
\newcommand{\post}{\mathrm{post}}
In this search $W$ and $Z$ bosons from the decay of the massive resonance are produced with a large transverse momentum relative to their mass and each boson is reconstructed as a single large-radius jet.  Boson-jet candidates are then identified by applying tagging requirements based on the reconstructed jet properties, as described below.

\subsection{Jet reconstruction}
Jets are formed by combining topological clusters \cite{topoclusters} reconstructed in the calorimeter system, which are calibrated in energy with the local calibration scheme \cite{lcw} and are considered massless. These topological clusters are combined into jets using the Cambridge--Aachen (C/A) algorithm~\cite{Dokshitzer:1997, cambridgeaachen1} implemented in FastJet~\cite{Cacciari:2011ma} with a radius parameter R = 1.2  with a radius parameter $R =1.2$.
The C/A algorithm iteratively replaces the nearest pair of elements (topological clusters or their combination) with their combination\footnote{Elements are combined by summing their four-momenta.} until all remaining pairs are separated by more than $R$, defining  the distance $\Delta R$ between elements as $(\Delta R)^2 = {(\Delta y)^2 + (\Delta\phi)^2}$ where $y$ is the rapidity. The jets are the elements remaining after this final stage of iteration, and the last pair of elements to be combined into a given jet are referred to here as the subjets of that jet.
Charged-particle tracks reconstructed in the tracking detector are matched to these calorimeter jets if they fall within the passive catchment area of the jet \cite{cacciari:catchment}, determined by representing each track by a collinear ``ghost'' constituent with negligible energy during jet reconstruction.  Only well-reconstructed tracks with \pt{} $\geq 500$ \mev{} and consistent with particles originating at the primary\footnote{The primary collision vertex is the reconstructed vertex with the greatest sum of associated track ${\pt}^2$.} collision vertex are considered.

The jets are then groomed to identify the pair of subjets associated with the $\Wboson\to q\bar{q}' $ or $\Zboson\to q\bar{q}$ decay, and to reduce the effect of pileup and other noise sources on the resolution. 
The grooming algorithm is a variant of the mass-drop filtering technique~\cite{bdrs},
which first examines the sequence of pairwise combinations used to reconstruct the jet in reverse order.  At each step the lower-mass subjet is discarded, and the higher-mass subjet is considered as the jet, continuing until a pair is found which satisfies \emph{mass-drop} and \emph{subjet momentum balance} criteria \brit{parameterised} by $\mu_\filt$ and $\sqrt{y_\filt}$, respectively.  Iteration stops when a pair of subjets is found for which each subjet mass $m_{(i)}$ satisfies $\mu \equiv {m_{(i)}}/{m_0} \leq \mu_\filt$, and for which 
\[ \sqrt{y} \equiv \text{min}({\pt}_{j_1},{\pt}_{j_2}) \frac{\Delta R_{(j_1,j_2)}}{m_0}  \geq \sqrt{y_\filt}\,,\]

where ${\pt}_{j_1}$ and ${\pt}_{j_2}$ are the transverse momenta of subjets $j_1$ and $j_2$ respectively, $\Delta R_{(j_1,j_2)}$ is the distance between subjets $j_1$ and $j_2$, and  $m_0$ is the mass of the parent jet.  

The subjet momentum-balance threshold $y_\filt$ and the mass-drop parameter $\mu_\filt$ used in this analysis, given in table~\ref{tab:rmdpars}, are chosen to stop the iteration when the two subjets corresponding to the \Wboson{} or \Zboson{} boson decay have been identified in simulated signal events. The best signal to background ratio for a given signal efficiency was obtained with  $\mu_\filt$ set to 1, hence no mass-drop requirement is applied in this analysis.

The selected pair of subjets is then filtered: the original topological cluster constituents of that pair of subjets are taken together and clustered using the C/A algorithm with a small radius parameter ($R_\mathrm{r}=0.3$), and all but the three ($n_\mathrm{r} = 3$) highest-\pt{} jets resulting from this reclustering of the subjets' constituents are discarded.  If there are fewer than three jets after the reclustering, all constituents are kept.  Those constituents that remain form the resulting filtered jet, which is further calibrated using energy- and $\eta$-dependent correction factors derived from simulation by applying a procedure similar to the one used in ref.~\cite{Aad:2013gja}. The calibrated four-momentum is used as the $W$ or $Z$ boson candidate's four-momentum in subsequent cuts and in reconstructing the heavy resonance candidate's mass in each selected event.

\begin{table}[!pht]
\caption{
  Parameters for the mass-drop filtering algorithm used to groom C/A jets. The choice of $\mu_\filt$ parameter corresponds to no mass-drop requirement being imposed in the grooming procedure.
\label{tab:rmdpars}
}
\begin{center}
\begin{tabular}{r|c}
\hline
\hline
Filtering parameter & Value \\
\hline
$\sqrt{y_\filt}$ & 0.2 \\
$\mu_\filt$ & 1 \\
$R_{\mathrm{r}}$ & 0.3 \\
$n_{\mathrm{r}}$ & 3 \\
\hline
\hline
\end{tabular}
\end{center}
\end{table}

\subsection{Boson jet tagging}
\begin{figure}[th!]
  \begin{center}
       \subfigure[]
              {
     \includegraphics[width=0.45\textwidth]{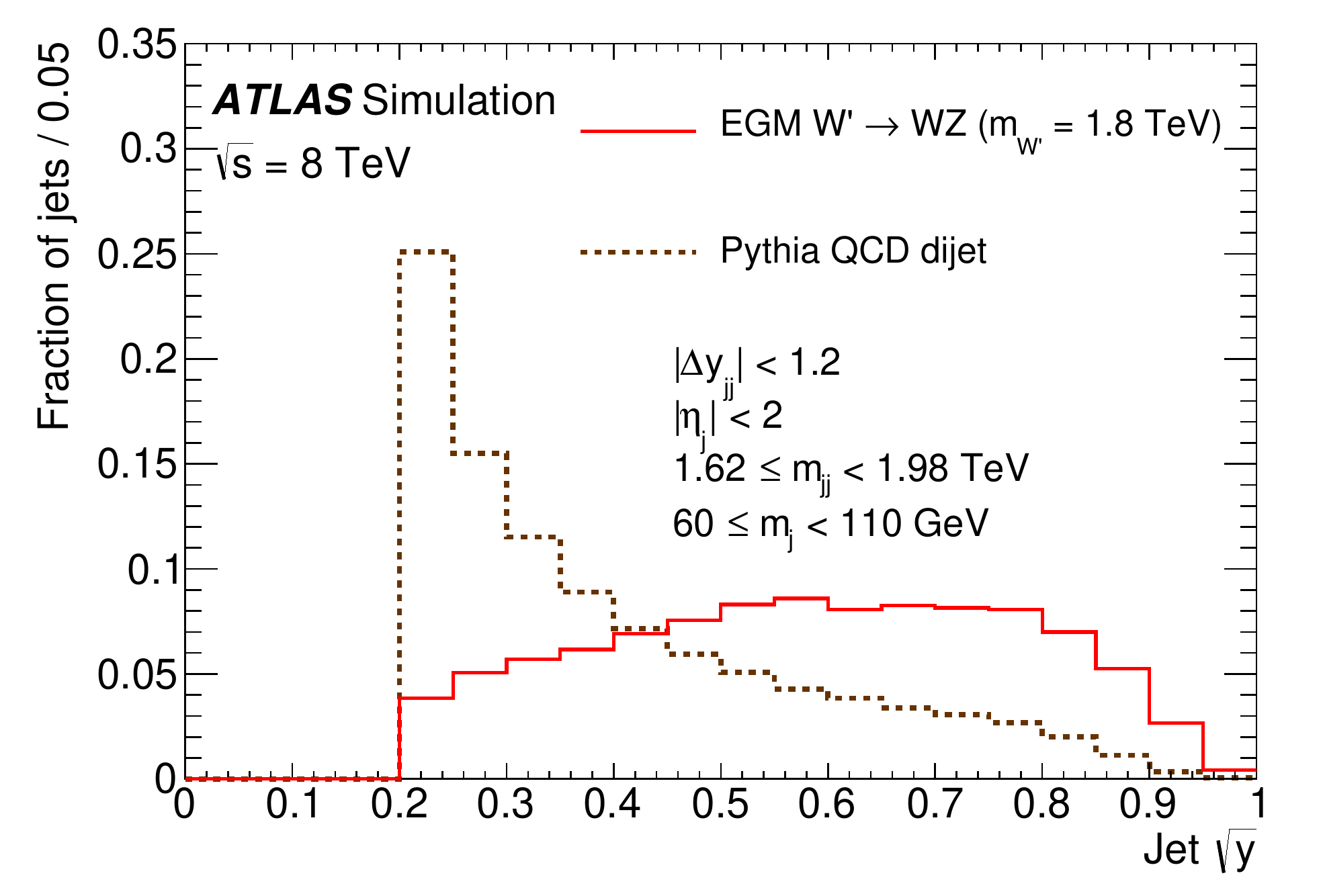}\label{fig:tagvars:y}
              }
    \subfigure[]
              {
     \includegraphics[width=0.45\textwidth]{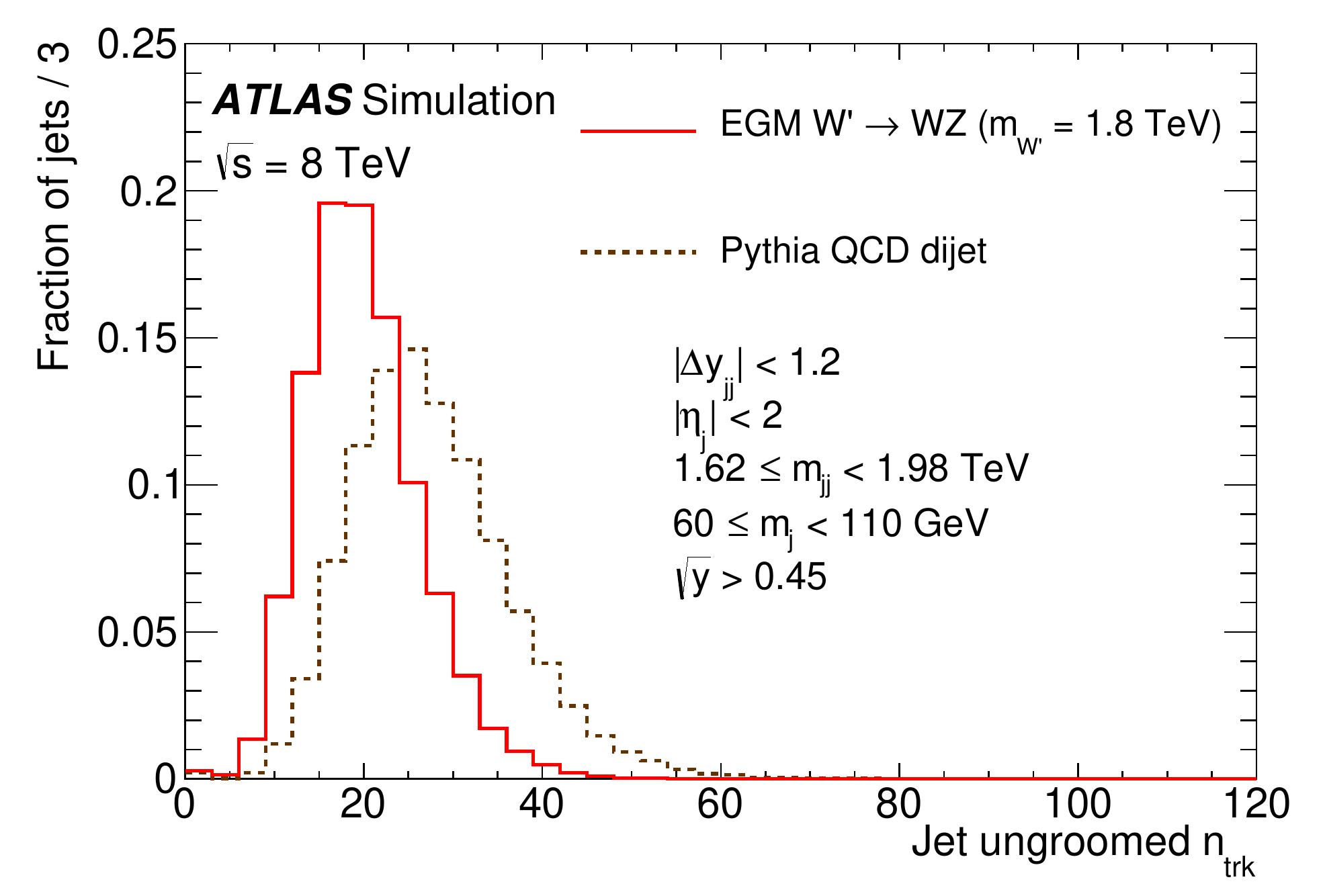}\label{fig:tagvars:trk}
              }
               \subfigure[]
              {
     \includegraphics[width=0.45\textwidth]{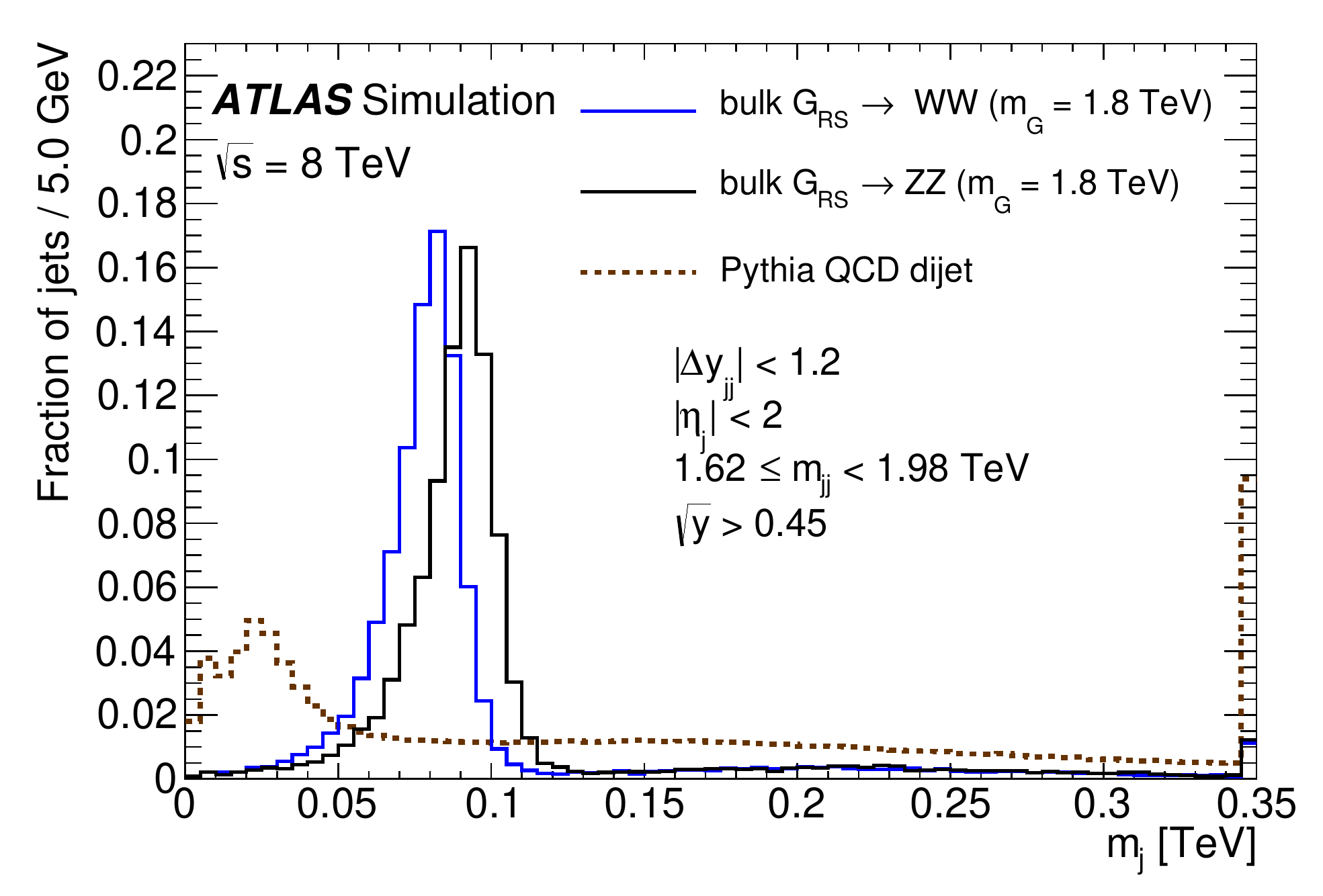}\label{fig:tagvars:mass}
              }

   \end{center}
\vspace{-20 pt}
     \caption{The distribution of the boson-tagging variables \subref{fig:tagvars:y} subjet momentum balance $\sqrt{y}$, \subref{fig:tagvars:trk} number of tracks $n_\text{trk}$ matched to the jet, and \subref{fig:tagvars:mass}  mass $m_j$ of the groomed jet, in simulated signal and background events.  
The signal and background distributions are \brit{normalised} to unit area, and the last bin of each histogram includes the fraction of events falling outside of the displayed range. Requirements are placed on the events used to ensure that the kinematics of the signal and background events are comparable. 
}
\label{fig:tagvars}
\end{figure}

The grooming algorithm rejects jets that do not satisfy the momentum balance and mass-drop criteria at any stage of iteration, and thus provides a small degree of discriminating power between jets from hadronically decaying bosons and those from QCD dijet production.  To improve the discrimination in this analysis, the remaining jets are also tagged with three additional boson tagging requirements.  First, a more stringent subjet momentum-balance criterion 
($\sqrt{y} \geq 0.45$) is applied to the pair of subjets identified by the filtering algorithm at the stopping point before the reclustering stage, since jets in QCD dijet events that survive grooming tend to have unbalanced subjet momenta characteristic of soft gluon radiation. Figure~\ref{fig:tagvars:y} shows the subjet momentum-balance distribution for jets in signal and QCD dijet background simulated events. 
Unlike jets from massive boson decays in which the hadron multiplicity is essentially independent of the jet \pt{}, energetic gluon jets are typically composed of
more hadrons, so the number of charged-particle tracks associated with the original, ungroomed jet is required to be small ($n_\text{trk} < 30$). Figure~\ref{fig:tagvars:trk} shows the number of tracks matched to jets for selected jets in signal and background simulated events.  The efficiency of this selection requirement must be corrected by a scale factor derived from data, as explained in section~\ref{sec:mcsig}.
Finally, a selection window is applied to the invariant mass of the filtered jet, $m_j$, 
since this quantity is expected to be small for jets in QCD dijet events and to reflect the boson mass for jets from hadronic boson decays.  The expected mass distribution of jets in \grav{} simulated events and the dijet background simulation is illustrated in figure~\ref{fig:tagvars:mass}.   
Narrow mass windows with a width of $26$ \GeV\ are chosen to optimise sensitivity to signal events and are centred at either 82.4 \gev{} or 92.8 \gev{}, where the mass distributions of the $W$ and $Z$ jets, respectively, peak in simulation. Each jet's mass must fall within either the $W$ or $Z$ mass window, consistent with the  $WZ$, $WW$ or $ZZ$ final state being studied. 

\section{Event selection}\label{sec:evsel}
High-mass resonances decaying to a pair of boosted vector bosons with subsequent hadronic decay are recognised as two large-radius massive jets with large momentum, typically balanced in \pt{}.  Events in this search must therefore first satisfy the high-\pt{} large-radius jet trigger, which is found to select over 99\% of C/A $R$=1.2 jets within $|\eta| < 2.0$ and with ungroomed \pt{} greater than 540 \GeV.   
Events are removed if they contain a prompt electron candidate with \et{} $> 20 \GeV$ in the regions $|\eta| < 1.37$ or $1.52 < |\eta| < 2.47$, or a prompt muon candidate with \pt{} $>20 \GeV$ in the region $|\eta| < 2.5$.  This requirement ensures that this analysis has no events in common with other diboson search analyses~\cite{atlas:llqq_2012,atlas:lnuqq_2012}.  Events with reconstructed missing transverse momentum exceeding 350 \GeV{} are also removed, as these are used in searches sensitive to diboson resonances with a $Z$ boson decaying to neutrinos~\cite{Aad:2013oja}.

\subsection{Event topology}
For events satisfying the requirements above, two C/A $R$=1.2 jets with \pt{} exceeding 20 GeV must be found and must pass the mass-drop filtering procedure.  The two jets with the highest transverse momentum must have $|\eta| < 2.0$ to ensure sufficient overlap with the inner tracking detector, since associated charged-particle tracks are used in the boson tagging requirements and for estimating systematic uncertainties.  In addition, a requirement on the rapidity difference between the two leading jets, $|y_1 - y_2| < 1.2$, is imposed to improve the sensitivity.  This rapidity difference is smaller for $s$-channel processes such as the $W'$ and $G_{RS}$ signal models than for the $t$-channel processes dominating the QCD dijet background.

The combined efficiency of these three cuts in $W'$ signal events is between 
72\% and 81\% 
depending on the resonance mass, for events from each signal sample in which the true diboson mass lies within 10\% of the nominal value.  For \grav{} signal events, the combined efficiency is between 
82\% and 87\%.  
The difference in the expected efficiencies between the $W'$ and  \grav{} signals is related to the different event topologies expected for spin-1 and spin-2 resonances affecting acceptance.

\begin{figure}[th!]
  \begin{center}
    \subfigure[]
              {
     \includegraphics[width=0.45\textwidth]{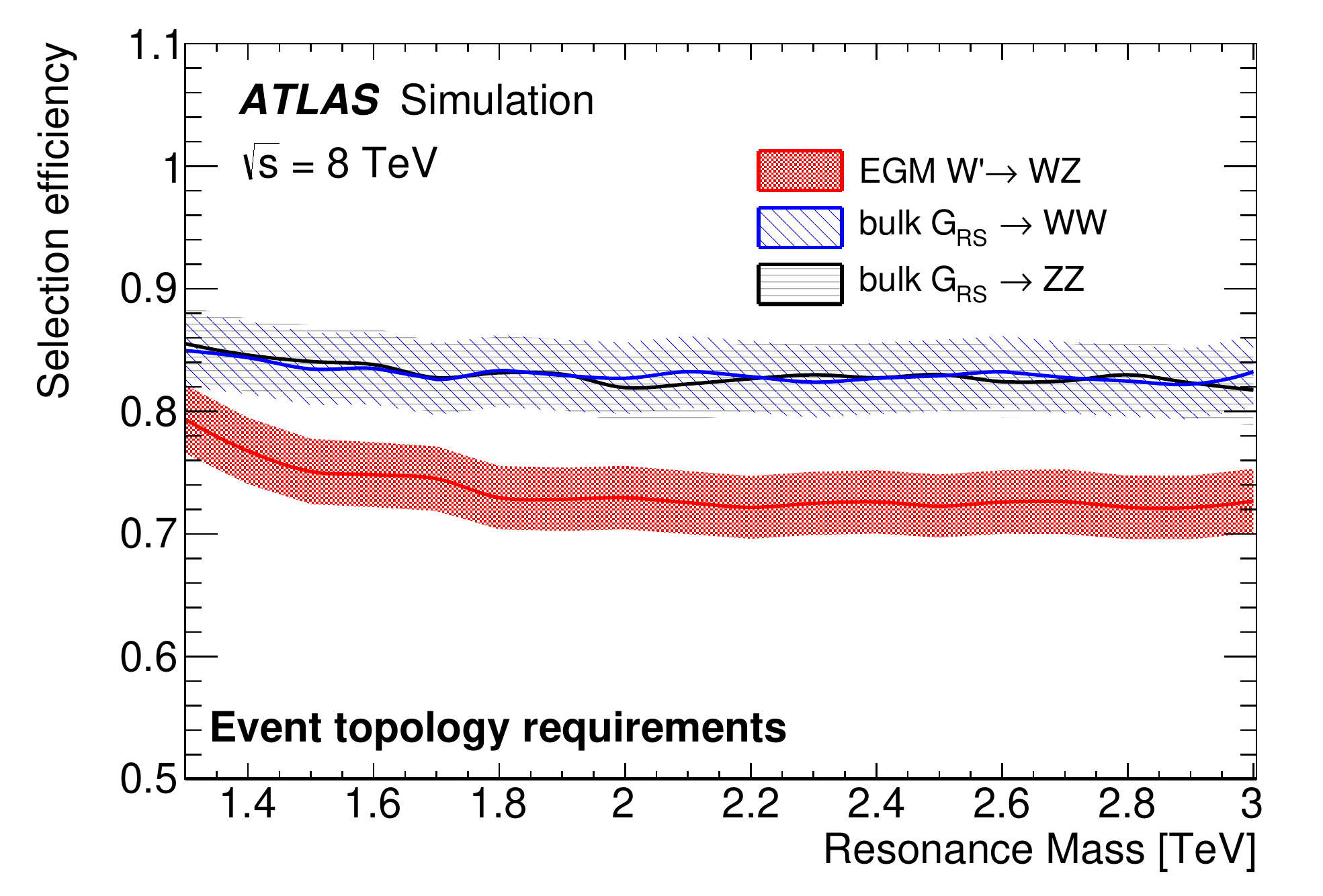}\label{fig:eff:evsel}
              }
    \subfigure[]
              {
     \includegraphics[width=0.45\textwidth]{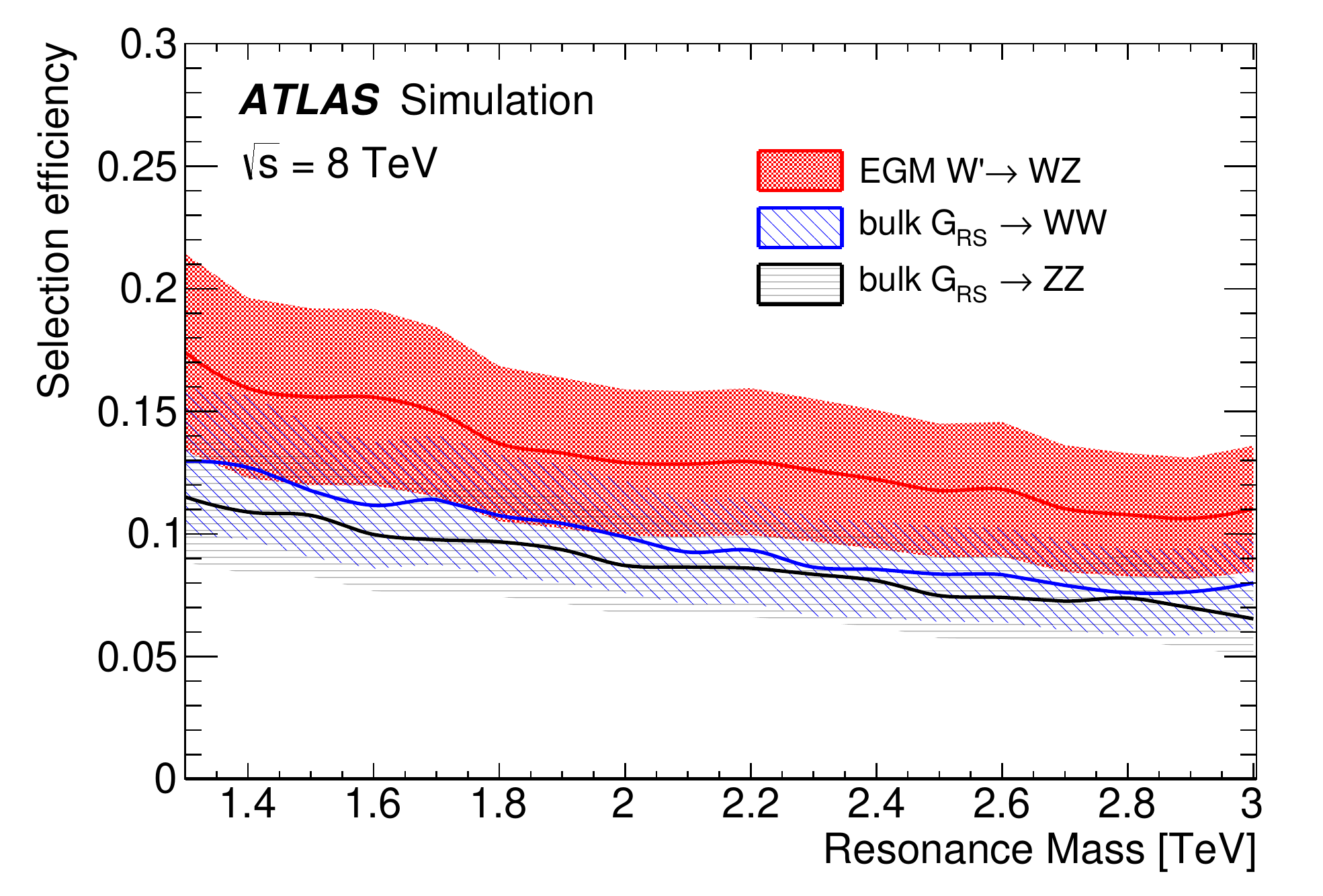}\label{fig:eff:tag}
              }
   \end{center}
\vspace{-20 pt}
     \caption{Event selection efficiencies as a function of the resonance masses for EGM $W'\to WZ$ and bulk $\grav \to WW$ and $ZZ$ for simulated events with resonance mass within 10\% of the nominal signal mass. In \subref{fig:eff:evsel}, the event topology requirements are applied to EGM $W'\to WZ$, $\grav \to WW$ and $\grav \to ZZ$ samples, while in \subref{fig:eff:tag}, the $WZ$, $WW$ and $ZZ$ boson tagging selections are also applied in the EGM $W'\to WZ$, $\grav \to WW$ and $\grav \to ZZ$ samples respectively and the efficiencies shown are corrected by the simulation--to--data scale factor. The width of the bands in each figure indicates both the statistical and systematic uncertainties. }
\label{fig:eff}
\end{figure}

A selection on the \pt{} asymmetry of the two leading jets, $({\pt}_1 - {\pt}_2)/({\pt}_1 + {\pt}_2) < 0.15$, is used to reject events where one of the jets is poorly measured or does not come from the primary $pp$ collision.  The signal selection efficiency of this cut exceeds 97\% in $W'$ signal samples and 90\% in the samples of \grav{} events. 
The difference in the expected efficiencies between the $W'$ and  \grav{} signals is related to their different production mechanisms. 
Figure~\ref{fig:eff:evsel} shows the selection efficiency of the event topology requirements for signal events with resonance mass within 10\% of the nominal signal mass for the  $W' \to WZ$, bulk $\grav{} \to WW$ and bulk $\grav{} \to ZZ$  benchmark models, with statistical and systematic uncertainties indicated by the width of the bands in the figure.  

\subsection{Boson tagging requirements}
The two jets with the highest transverse momentum each must satisfy the three boson tagging requirements discussed in section~\ref{sec:bosid}:
$\sqrt{y} \geq 0.45$,  $n_\text{trk} < 30$, and $|m_j - m_V| < 13 \GeV$, where $m_V$ is the peak value of the reconstructed $W$ or $Z$ boson mass distribution. For the \wpwz\ search, this final cut sets $m_V$ equal to the peak reconstructed $W$ boson mass when applied to the lower mass jet, and to the peak reconstructed $Z$ boson mass when applied to the higher mass jet.

The expected efficiency of these boson tagging cuts applied to signal events is evaluated using the MC signal samples described in section~\ref{sec:sim}. For signal events passing event topology requirements on the mass-drop filtering, $\eta$, the rapidity difference, and the \pt{} asymmetry, the average efficiency of the tagging cuts for each of the two leading-\pt{} filtered jets is approximately the same in the $\grav \to WW$ and $\grav \to ZZ$ samples, and ranges from 44.0\% in the $m_{\grav}=1.2 \TeV$ sample to 33.9\% for the $m_{\grav}=3.0 \TeV$ sample.
Figure~\ref{fig:eff:tag} shows the selection efficiency of the event selection and tagging requirements for signal events with resonance mass within 10\% of the nominal signal mass for the $W' \to WZ$ and bulk $\grav{} \to WW$ and bulk $\grav{} \to ZZ$ benchmark models, with both statistical and systematic uncertainties included in the error band.
The average background selection efficiency of the tagger for each of the two leading-\pt{} filtered jets in simulated QCD dijet events satisfying the same event selection requirements ranges from 1.2\% for events with dijet masses between 1.08 \TeV{} and 1.32 \TeV{}, to 0.6\% for events with dijet masses between 2.7 \TeV{} and 3.3 \TeV{}.

\subsection{Dijet mass requirement}
The invariant mass calculated from the two leading jets must exceed 1.05 \TeV{}.  This requirement restricts the analysis of the dijet mass distribution to regions where the trigger is fully efficient for boson-tagged jets, so that the trigger efficiency does not affect its shape.

\section{Background model}
\label{sec:bg}

The search for high-mass diboson resonances is carried out by looking for resonance structures on a smoothly falling dijet invariant mass spectrum, empirically characterised by the function
\begin{equation}
\label{eqn:dijet1}
 \frac{\mathrm{d}n}{\mathrm{d}x} =  p_1 (1-x)^{p_2 + \xi p_3} x^{p_3},
\end{equation}
where $x = m_{jj}/\sqrt{s}$, and $m_{jj}$ is the dijet invariant mass, $p_1$ is a normalisation factor, $p_2$ and $p_3$ are dimensionless shape parameters, and $\xi$ is a dimensionless constant chosen after fitting to minimise the correlations between $p_2$ and $p_3$. 
A maximum-likelihood fit, with parameters $p_1$, $p_2$ and $p_3$ free to float,  is performed in the range  $1.05 \tev < \mjj < 3.55 \tev$, where the lower limit is dictated by the point where the trigger is fully efficient for tagged jets and the upper limit is set to be in a region where the data and the background estimated by the fit are well below one event per bin for the tagged distributions. 
The likelihood is defined in terms of events binned in 100-\gev{}-wide bins in $m_{jj}$ as
\begin{equation}
\label{eqn:likelihood}
\mathcal{L}= \prod_{i} \frac{\lambda_{i}^{n_{i}}e^{-\lambda_{i}}}{n_i!},
\end{equation}
where $n_{i}$ is the number of events observed in the $i^{\mathrm{th}}$ $m_{jj}$ bin and $\lambda_i$ is the background expectation for the same bin. 

\begin{figure}[th!]
  \begin{center}
 \subfigure[]
              {
     \includegraphics[width=0.45\textwidth]{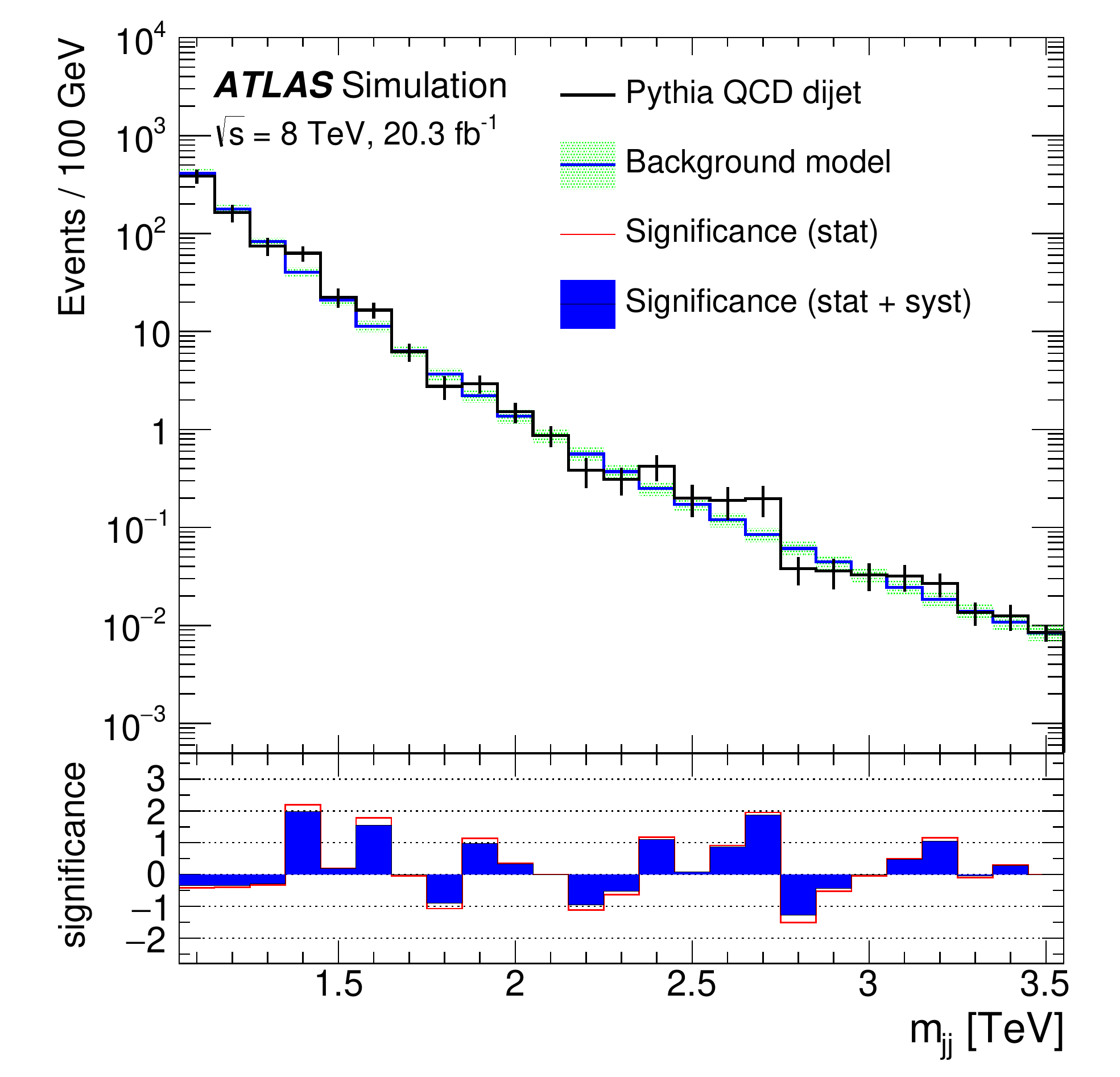}              
  }
 \subfigure[]
              {
	 \includegraphics[width=0.45\textwidth]{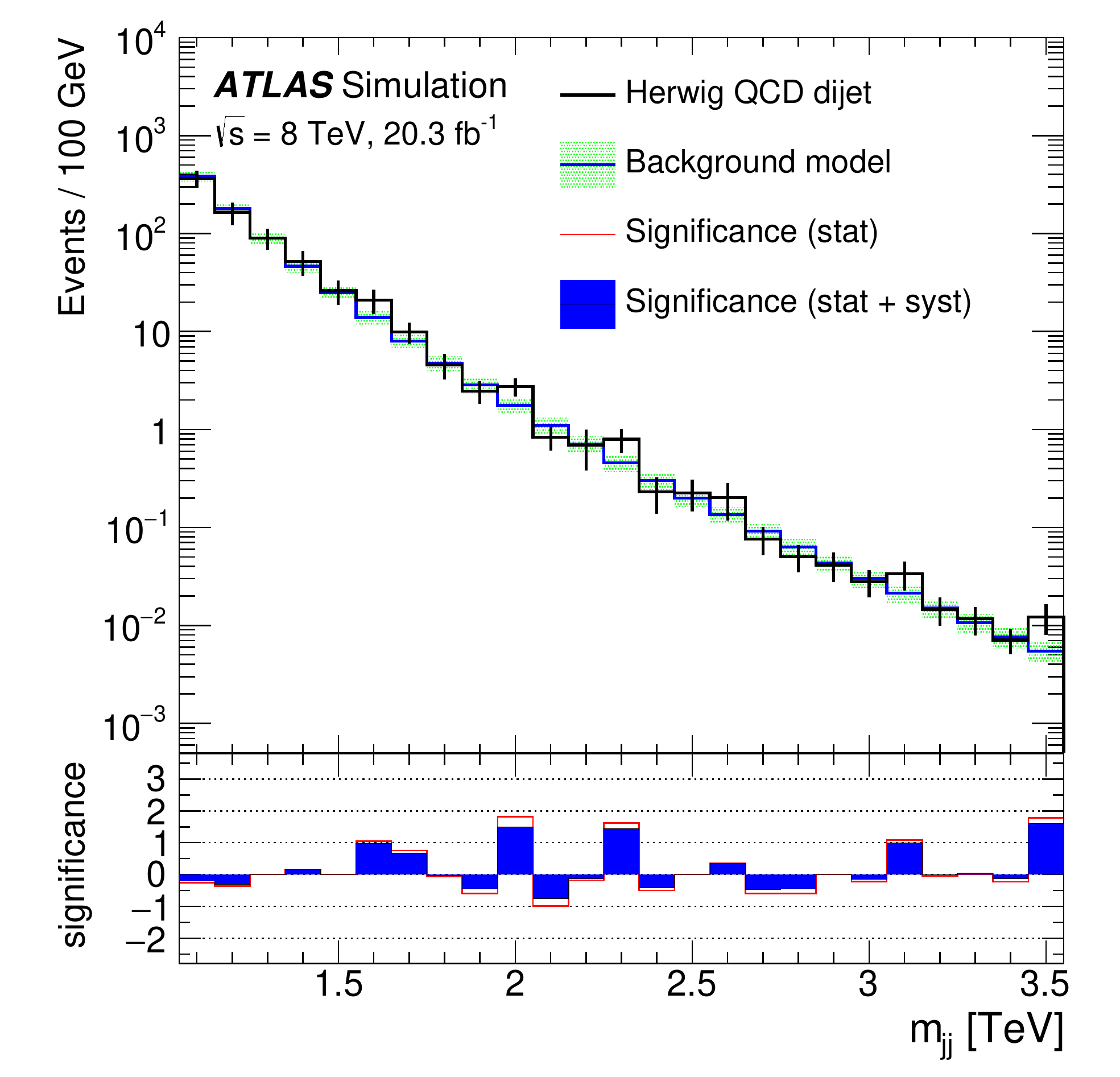}          
}
\end{center}
\vspace{-20 pt}
     \caption{Fits of the background model to the dijet mass ($m_{jj}$) distributions in (a) \pythia{} 8 and (b) \herwigpp{} simulated background events that have passed all event selection and tagging requirements.  The events are reweighted in both cases to correctly reproduce the leading-jet \pt{} distribution for untagged events, and the simulated data samples were scaled to correspond to a luminosity of 20.3 fb$^{-1}$.
The significance shown in the inset for each bin is calculated using the statistical errors of the simulated data.}
\label{fig:djmfit_mc}
\end{figure}

\begin{figure}[th!]
  \begin{center}
  \subfigure[]
              {
     \includegraphics[width=0.45\textwidth]{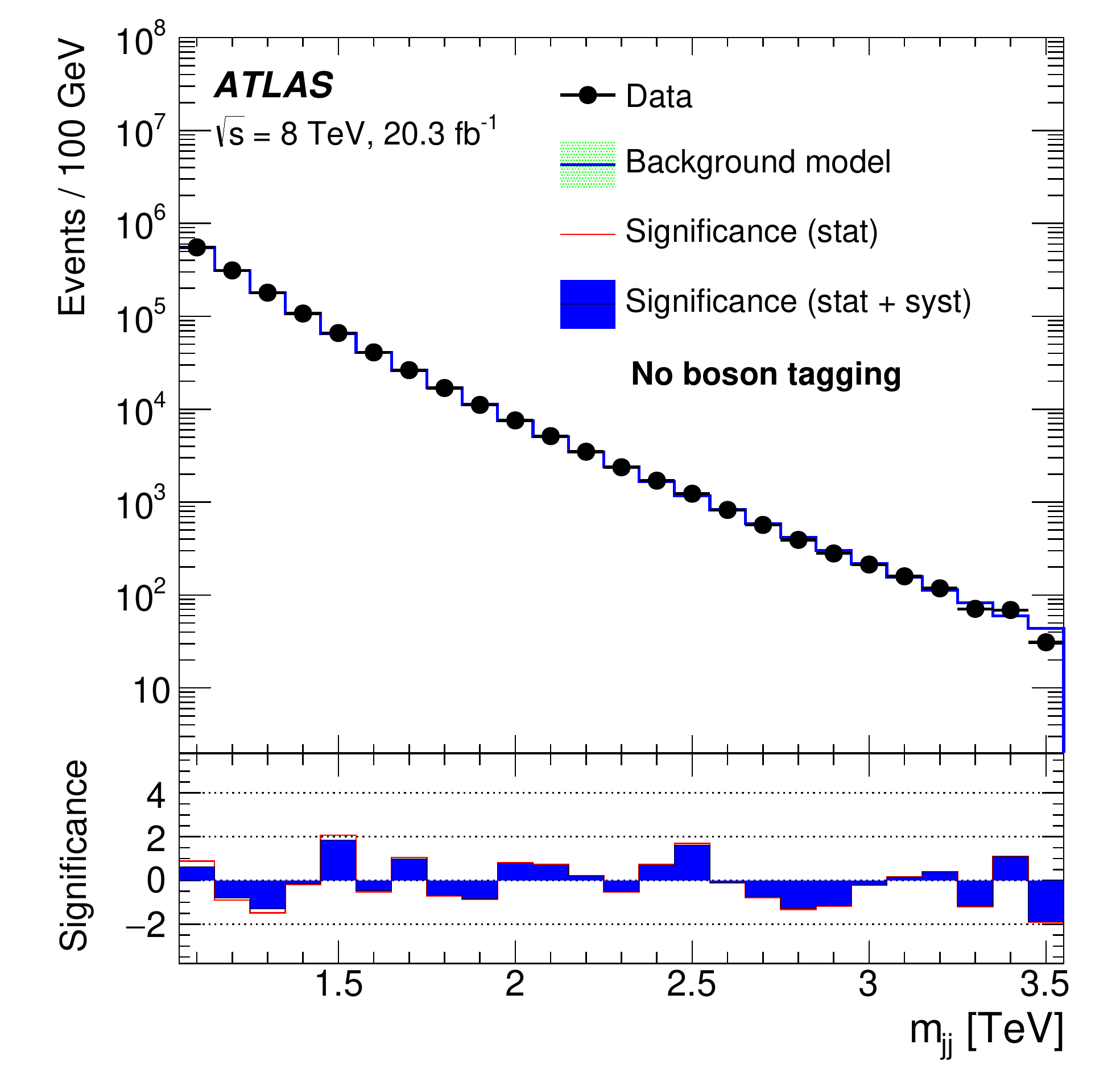}\label{fig:djm_fitsb:notag}
              }
   \subfigure[]
		{
     \includegraphics[width=0.45\textwidth]{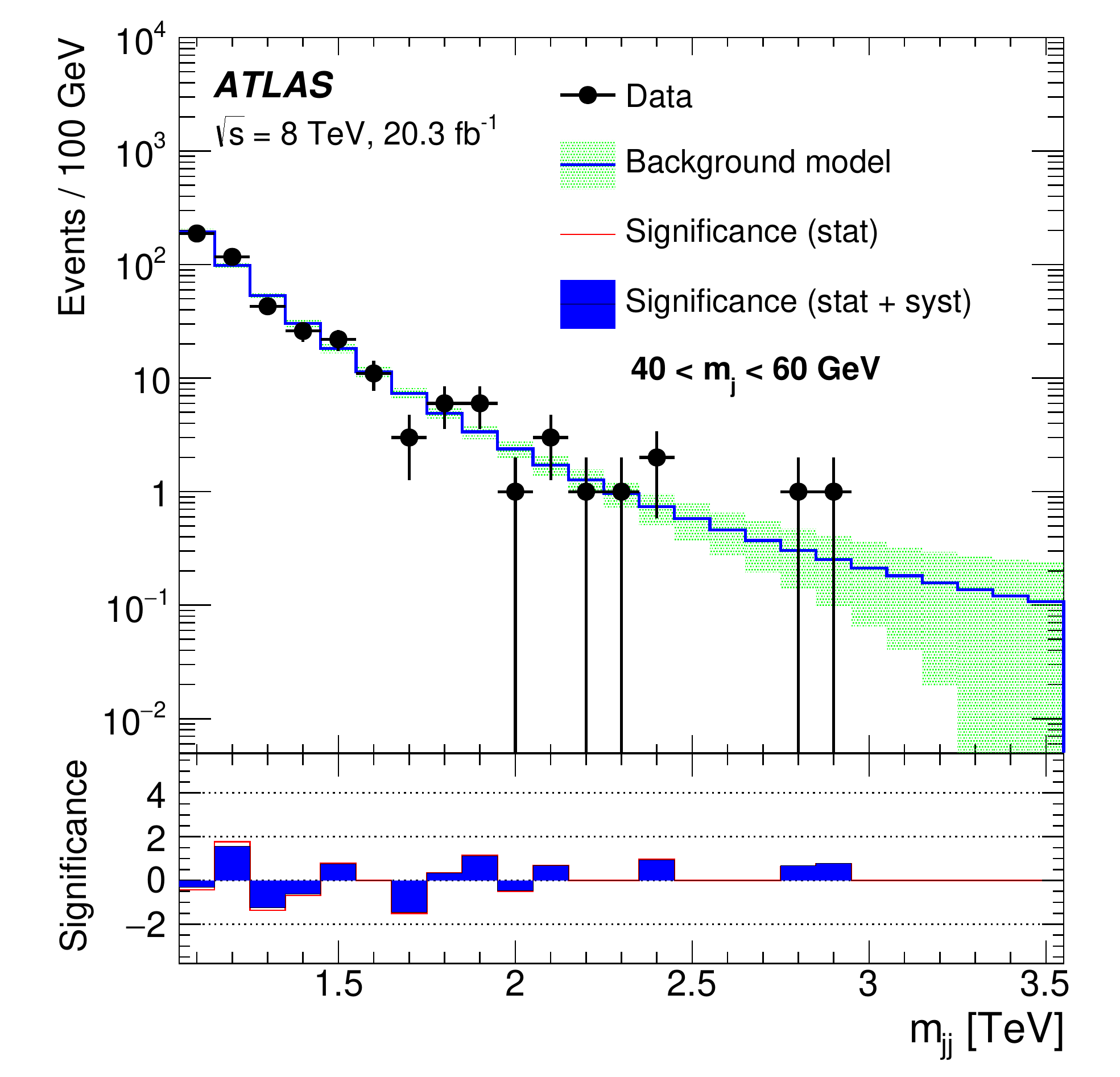}\label{fig:djm_fitsb:lowmass}              
		}	
\end{center}
\vspace{-20 pt}
     \caption{Fits of the background model to the dijet mass ($m_{jj}$) distributions in data events (a) before boson tagging, and (b) where both jets pass all tagging requirements except for the $m_j$ requirement, and instead satisfy  $40 < m_j \leq 60 \GeV$.  }
\label{fig:djmfit_sidebands}
\end{figure}

The functional form in eq.~\ref{eqn:dijet1} is tested for compatibility with distributions similar to the expected background by applying it to simulated background events and to several sidebands in the data.  Figure~\ref{fig:djmfit_mc} shows fits to the \herwigpp{} and \pythia{} simulated dijet events that pass the full event selection and tagging requirements on both jets, where the predictions from these leading-order generators are corrected by reweighting the untagged leading-jet \pt{} distributions to match the untagged distribution in data.
Figure~\ref{fig:djmfit_sidebands} shows the results of fitting the dijet mass spectrum before tagging, and in otherwise tagged events where both the leading and subleading jet have masses falling below the boson-tagging mass windows, in the range $40 < m_j < 60 \GeV$.  For the data selected before boson tagging, the trigger efficiency as a function of the dijet mass is taken into account in the fit, because in the untagged jet sample the trigger is not fully efficient in the first dijet mass bin displayed. 
The fitted background functions in figures~\ref{fig:djmfit_mc} and \ref{fig:djmfit_sidebands} are integrated over the same bins used to display the data, and labelled ``background model'' in the figures.  The size of the shaded band reflects the uncertainties of the fit parameters.  In figure~\ref{fig:djm_fitsb:notag} the background model is shown, but the size of the uncertainty band is too small to be seen. The lower insets in the figures show the significance, defined as the signed $z$-value of the difference between the distribution being modelled and the background model's prediction \cite{Choudalkis:2012}. The significance with respect to the maximum-likelihood expectation is displayed in red, and the significance when taking the uncertainties on the fit parameters into account is shown in blue.

Table~\ref{tab:djmfit_checks} summarises the results of these fits, as well as fits to data where one jet mass falls in the low-mass sideband ($40 < m_j \leq 60 \GeV$) and the other falls in a high-mass sideband from  $110  < m_{j} < 140 \gev{}$, and where both jet masses fall in the high-mass sideband.  
 The dijet mass distribution of the simulated background and of each of these background-dominated selections are well-described by the functional form in eq.~(\ref{eqn:dijet1}). 

\begin{table}[!pht]
\caption{
  Goodness-of-fit for maximum-likelihood fits of the background model to the dijet mass distribution in simulated events, and in selected mass sidebands from data events where at least one of the leading and subleading jet fails the jet mass selection.  One-sided $\chi^2$ probabilities are displayed; for the three data sideband fits, these probabilities were calibrated using pseudo-experiments to avoid biases due to empty bins.
\label{tab:djmfit_checks}
}
\begin{center}
\begin{tabular}{p{9cm}|c|c}
\hline
\hline
Sample & $\chi^2/\mathrm{nDOF}$ & Probability \\
\hline
\pythia{} dijet events& 24.6/22 & 0.31 \\
\herwigpp{} dijet events& 15.9/22 & 0.82 \\\hline
Data with $110  < m_{j1} \leq 140 \gev{}$ and  $40  < m_{j2} \leq 60 \gev{}$ & 12.1/11 & 0.79 \\
Data with $40  < m_{j} \leq 60 \gev{}$  for both jets  & 19.8/13 & 0.56 \\
Data with $110  < m_{j} \leq 140 \gev{}$ for both jets & 5.0/6 & 0.91 \\
\hline
\hline
\end{tabular}
\end{center}
\end{table}

\section{Systematic uncertainties}
\label{sec:mcsig}

The uncertainty on the background expectation is determined by the fitting procedure, which assumes a smoothly falling $m_{jj}$ distribution.  Possible uncertainties due to the background model were assessed by investigating several alternative families of parametrisations, and by considering signal plus background fits of the chosen function
to simulations of the dominant background as well as sidebands and control regions of data in which a signal contribution is expected to be negligible. These effects
 were estimated to be no more than 25\% of the statistical uncertainty at any mass in the search region. The effect of the uncertainty on the trigger efficiency, the variations of the selection efficiencies as a function of the kinematic properties of the background, and the composition of the background were also studied and were found to be well-covered by the uncertainties from the fit.

Systematic uncertainties on the shape of the $m_{jj}$ distribution and the normalisation of the $W'$ and \grav{} signal are expressed as nuisance parameters with specified probability distribution functions (pdfs).  The overall normalisation is a product of scale factors, each corresponding to an identified nuisance parameter.  If the shape is affected by a given nuisance parameter, the systematic 
change is included when the signal distribution is generated.  If the nuisance parameter does not affect the shape, but only affects the 
normalisation, the distribution is simply scaled.

The jet \pt\ scale $\alpha_{\mathrm{PT}} $ is defined as a multiplicative factor to the jet \pt{} in simulation, \(  \pt = \alpha_\mathrm{PT}\  \pt^\mathrm{MC} \).
Following the technique used in ref.~\cite{Aad:2013gja}, the systematic uncertainty on $\alpha_{\mathrm{PT}}$ is assessed by applying the jet reconstruction and filtering algorithms to inner-detector track constituents, which are treated as massless, and matching these track jets to the calorimeter jets.  The ratio of the matched track jet's \pt{} to the calorimeter jet's \pt{} as a function of several kinematic variables is compared in simulation and data and found to be consistent within 2\%.  Hence, the pdf used for $\alpha_\mathrm{PT}$ is a Gaussian with a mean of one and a standard deviation of 0.02.  Similar methods are used to determine the {pdf}s for the scale uncertainties in jet mass $m_j$  and momentum balance $\sqrt{y}$.

Mismodelling the jet \pt\ resolution can change the reconstructed width of a diboson resonance.  The jet \pt\ resolution in the simulation is 5\% in this kinematic region, and a 20\% systematic uncertainty on this resolution is implemented by applying a multiplicative smearing factor, $r_\mathrm{E}$, to the $p_\mathrm{T}$ of each reconstructed jet with a mean value of unity and a width $|\sigma_{r_E}|$.  The nuisance parameter $\sigma_{r_E}$ represents the uncertainty in the \pt{} resolution of reconstructed jets and is assumed to have a Gaussian {pdf} with a mean of zero and a standard deviation \( 0.05 \times \sqrt{1.2^2 - 1^2} \). 

The pdfs for the uncertainties in jet-mass resolution and momentum-balance resolution, listed in table~\ref{tab:systsummary}, are similarly constructed.

The $n_{\text{trk}}$ variable is not modelled sufficiently well in simulation \cite{quarkgluon7tev}, so it is necessary to apply a scale factor to the simulated signal to correct the selection efficiency 
of the $n_{\mathrm{trk}}$ requirement.  A scale factor of $0.90 \pm 0.08$ is derived from the ratio of the selection efficiency of this cut in a data control region enriched with $W/Z$ + jets events, 
where a high-\pt{} $W$ or $Z$ boson decays hadronically, to the selection efficiency of this cut in simulation. 
The data control region is defined by selecting events in the kinematic range where the  
jet trigger used in the search is fully efficient, and where only the leading-\pt{} jet passes the tagging requirement on $\sqrt{y}$.  
Fits to the jet mass spectrum of the leading-\pt{} jet determine the number of hadronically decaying $W$ and $Z$ bosons reconstructed as a single jet that pass the $n_{\mathrm{trk}}$ requirement as a function of the selection criteria on $n_{\mathrm{trk}}$.  
The dominant uncertainty on these yields is the mismodelling of the jet mass spectrum for non-$W$ or non-$Z$ jets, and is evaluated by comparing the yields obtained when using two different background models in the fit.  
The resulting scale factor is $0.90 \pm 0.08$. Since the $n_{\mathrm{trk}}$ requirement is applied twice per event in the selections used in the search,  a scale factor of $0.8$ is applied per selected signal event, with an associated uncertainty of 20\%. 
	         
A 5\% uncertainty on the signal efficiency due to uncertainties on the parton shower and hadronisation model is also included. The uncertainty is estimated by comparing the selection efficiencies obtained in simulated signal samples generated and showered with \pythia{} 8 to the selection efficiencies obtained in samples generated with \pythia{} 8 and showered with \herwigpp. An additional 3.5\% uncertainty on the signal acceptance due to uncertainties on the PDFs is considered. This uncertainty is estimated according to the PDF4LHC recommendations~\cite{Botje:2011sn}.  
	 
Table~\ref{tab:systsummary} summarises the systematic uncertainties affecting the signal shape and the {pdf} constraining the associated nuisance parameter.
The largest uncertainty on the shape of the reconstructed signal is due to the jet \pt\ scale and resolution; the uncertainty in the scale introduces 
an uncertainty on the scale of the mass of the reconstructed resonant signal, and the resolution introduces an uncertainty of the width. The jet mass scale uncertainty 
also has an effect on the scale of the reconstructed mass, but this effect is less significant. Table~\ref{tab:systsummary_norm} summarises  
the systematic uncertainties affecting the signal normalisation. The jet mass scale uncertainty affects both the shape and the normalisation.  

\begin{table}[!pht]
\caption{
Summary of the systematic uncertainties affecting the shape of the signal dijet mass distribution and their corresponding models. \(G(x|\mu,\sigma)\) in the table denotes a Gaussian distribution for the variable \(x\) with mean \(\mu\) and standard deviation \(\sigma\).
\label{tab:systsummary}
}
\begin{center}
\begin{tabular}{r|cc}
\hline
\hline
Source & Uncertainty & Constraining {pdf} \\
\hline
Jet \pt\ scale & 2\% & $G(\alpha_\mathrm{PT}|1, 0.02)$ \\
Jet \pt\ resolution & 20\% & $G(\sigma_{r_E}|\,0, 0.05 \times \sqrt{1.2^2 - 1^2}) $ \\
Jet mass scale & 3\% &$G(\alpha_\mathrm{m}|1, 0.03)$ \\
\hline
\hline
\end{tabular}
\end{center}
\end{table}

\begin{table}[!pht]
\caption{
Summary of the systematic uncertainties affecting the signal normalisation and their impact on the signal.
\label{tab:systsummary_norm}
}

\begin{center}
\begin{tabular}{r|c}
\hline
\hline
Source & Normalisation uncertainty \\
\hline
Efficiency of the track-multiplicity cut & 20.0\% \\
Jet mass scale  & 5.0\% \\
Jet mass resolution  &  5.5\% \\
Subjet momentum-balance scale  &  3.5\% \\
Subjet momentum-balance resolution & 2.0\% \\
Parton shower model  & 5.0\% \\
Parton distribution functions  & 3.5\% \\
Luminosity &  2.8\%  \\

\hline
\hline
\end{tabular}
\end{center}
\end{table}

\section{Results}\label{sec:result}
\subsection{Background fit to data}
\begin{figure}[th!]
  \begin{center}
    \subfigure[]
              {
     \includegraphics[width=0.45\textwidth]{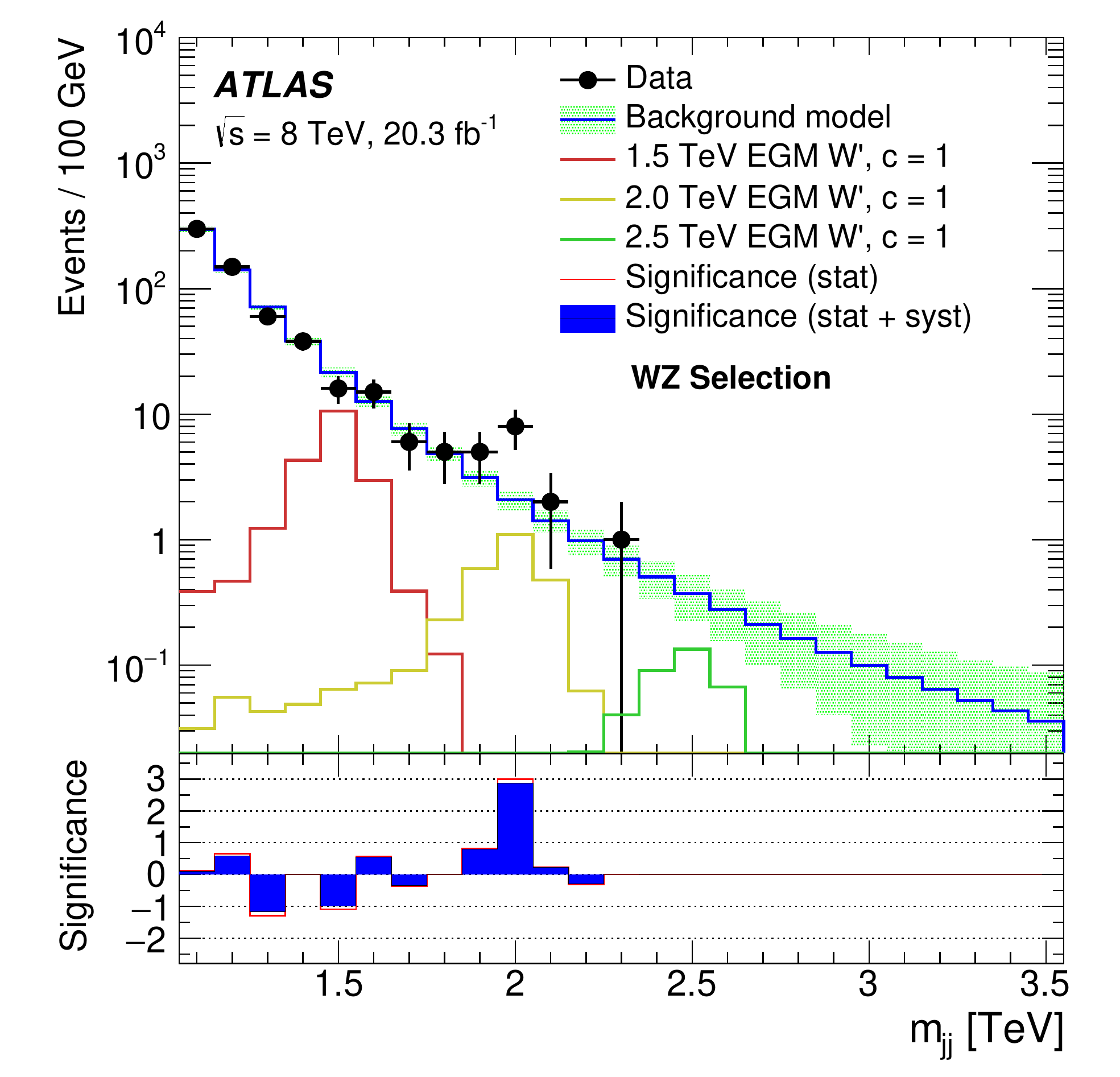}\label{fig:djm_fit:wz}
              }
    \subfigure[]
              {
     \includegraphics[width=0.45\textwidth]{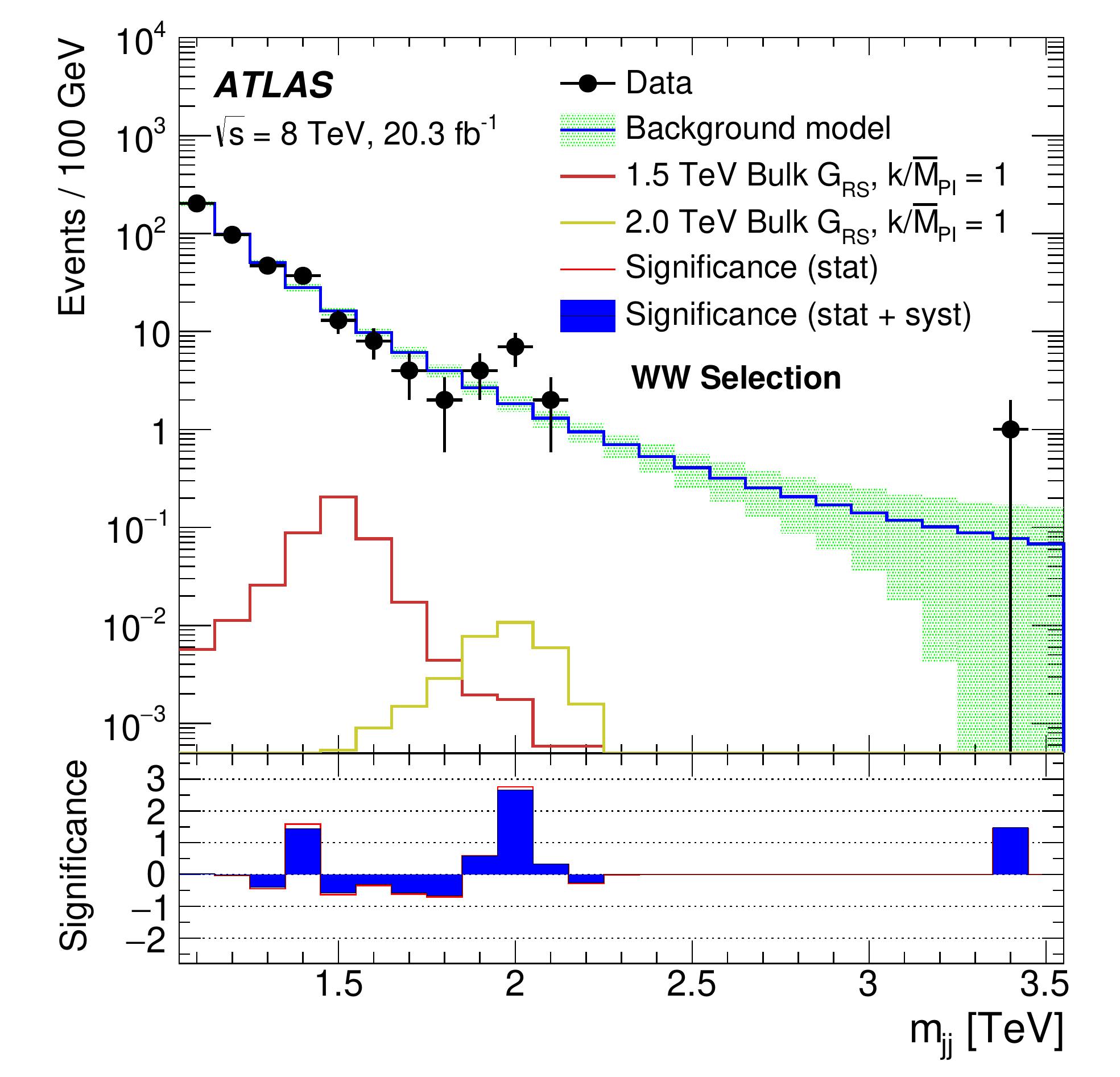}\label{fig:djm_fit:ww}
              }
     \subfigure[]
              {
     \includegraphics[width=0.45\textwidth]{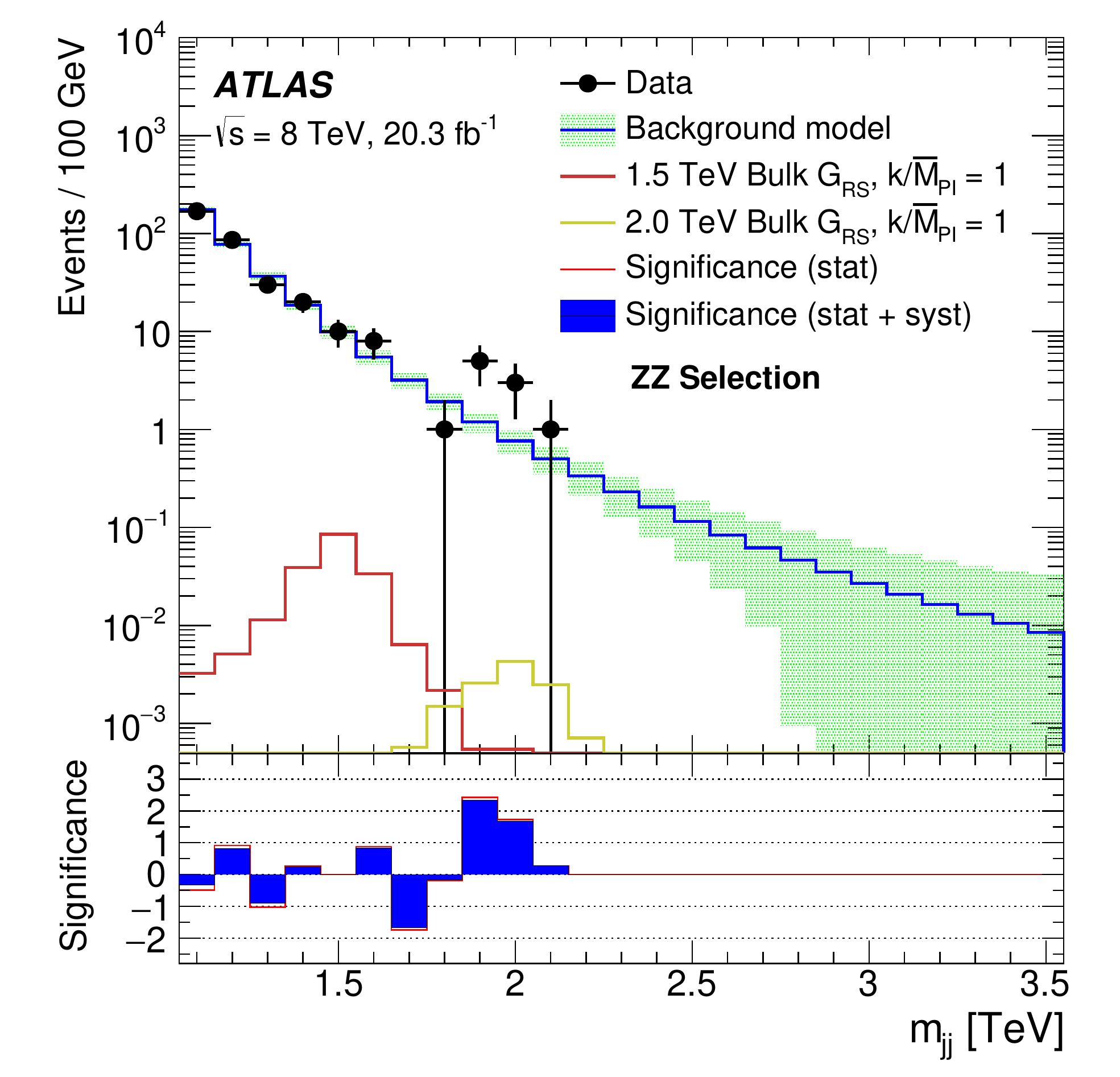}\label{fig:djm_fit:zz}
              }
	\subfigure[]
              {
     \includegraphics[width=0.45\textwidth]{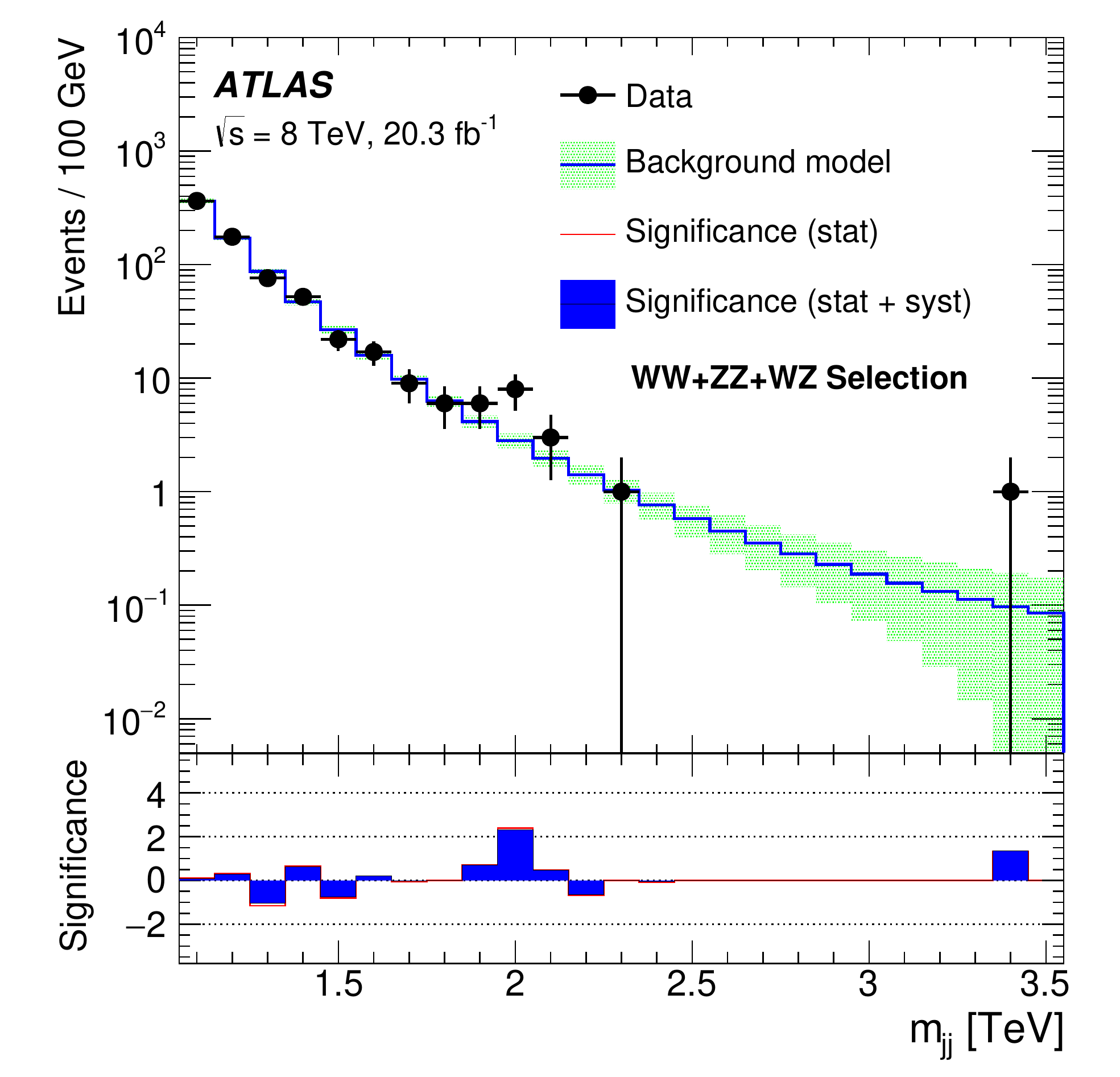}\label{fig:djm_fit:inc}
              }
\end{center}
\vspace{-20 pt}
     \caption{Background-only fits to the dijet mass ($m_{jj}$) distributions in data \subref{fig:djm_fit:wz}  after tagging with the $WZ$ selection, \subref{fig:djm_fit:ww} after tagging with the $WW$ selection, \subref{fig:djm_fit:zz} after tagging with the $ZZ$ selection,
 and \subref{fig:djm_fit:inc} for events passing any of the three tagging selections.
The significance shown in the inset for each bin is the difference between the data and the fit in units of the uncertainty on this difference.
The significance with respect to the maximum-likelihood expectation is displayed in red, and the significance when taking the uncertainties on the fit parameters into account is shown in blue.
The spectra in the three signal regions are compared to the signals expected for an EGM $W'$ with $m_{W'}$ = 1.5, 2.0, or 2.5 \tev{} or to an RS graviton with $m_{\grav}$ = 1.5 or 2.0 \tev{}.}
\label{fig:djm_fit}
\end{figure}

The fitting procedure is applied to the data after $WZ$, $WW$ and $ZZ$ selection, and 
 the results are shown in figure~\ref{fig:djm_fit}.  In this figure, the fitted background functions, labelled ``background model,'' are again integrated over the same bins used to display the data,  and the size of the shaded band reflects the uncertainties on the parameters propagated to show the uncertainty on the expectation from the fit. 
Figure~\ref{fig:djm_fit} also displays
the fitted dijet mass distribution of events passing any of the three tagging selections. 
The lower panels in the figure show the significance of the difference between data and the expectation in each bin.
 Table~\ref{tab:djm_pars} gives the fitted values of the parameters for the data selected before tagging, displayed in figure~\ref{fig:djm_fitsb:notag}, and after the $WZ$, $WW$ and $ZZ$ selections, as well as the number of events observed.

\begin{table}[!pht]
\caption{
  Number of observed events and parameters from the background-only fits to the dijet mass spectrum for each tagging selection. The parameter $\xi$ is a 
  constant chosen after the fit to minimise the correlation between the fitted parameters $p_2$ and $p_3$. 
\label{tab:djm_pars}
}
\begin{center}
\begin{tabular}{c|c|c|c|c}
\hline
\hline
Parameter & Before tagging & $WZ$ & $WW$ & $ZZ$ \\
\hline
 $\xi$ & 4.3 & 3.8 & 4.2 &  4.5 \\
$p_2$ & $ 30.95 \pm  0.03 $ & $31.0 \pm 1.4$ & $32.5 \pm 1.5$ &  $39.5 \pm 2.0$ \\
$p_3$ & $-5.54 \pm 0.03$ & $-9.1 \pm 1.5 $ & $ -9.4 \pm 1.6 $ & $  -9.5 \pm 2.3$ \\
\hline
Observed events & 1335762 & 604 & 425 & 333 \\
\hline
\hline
\end{tabular}
\end{center}
\end{table}

The dijet mass distributions after all three tagging selections are well-described by the background model over the entire mass range explored, with the exception of a few bins near $m_{jj} = 2\tev$ which contain more events than predicted by the background model. 
Approximately 20\% of the events selected by either the $WW$, $WZ$, or $ZZ$ selection are shared among all three signal regions. The fraction of events common to the $WZ$ and the $WW$ or the $WZ$ and the $ZZ$ selections are 49\% and 43\% respectively. 
After requiring that $m_{jj} > 1.75 \TeV$, 5 out of 25 events are common to all three signal regions. The statistical interpretation of these dijet mass distributions is discussed in the following section. 

\subsection{Statistical analysis}\label{sec:sa}

A frequentist analysis is used to interpret the data. For each of the two benchmark models under test, the parameter of interest in the statistical analysis is the signal strength, $\mu$, defined as a scale factor on the total number of signal events predicted by the model. Thus, the background-only hypothesis corresponds to $\mu=0$, and the hypothesis of a signal-plus-background model corresponds to $\mu=1$. A test statistic, $\lambda(\mu)$, based on the profile likelihood ratio~\cite{Cowan:2010js} is used to test these  models. The test statistic is designed to extract the information on $\mu$ from a maximum-likelihood fit of the signal-plus-background model to the data. 

The likelihood model for the observation is
\begin{equation}
\label{eqn:lhood}
  {\cal L} = \prod_i P_{\mathrm{pois}}( \nobsi | \nexpi ) \times G(\alpha_\mathrm{PT}) \times G(\alpha_\mathrm{m}) \times G(\sigma_{r_E}) \times {\cal N}(\mathbf{\theta}) 
\end{equation}
where $P_{\mathrm{pois}}( \nobsi | \nexpi )$ is the Poisson probability to observe $\nobsi$ events
if $\nexpi$ events are expected, $G(\alpha_\mathrm{p_T})$, $ G(\alpha_\mathrm{m})$, and $G(\sigma_{r_E})$, are the {pdf}s of the nuisance parameters modelling the systematic uncertainties related to the shape of the signal, and ${\cal N}$ is a log-normal distribution for the nuisance parameters, $\mathbf{\theta}$,  modelling the systematic uncertainty on the signal normalisation. 
The expected number of events is the bin-wise sum of the events expected for the
signal and background:
  $\nexpvec = \nsigvec + \nbgvec$.
The number of expected background events in dijet mass bin $i$, \nbgi,
is obtained by integrating $\mathrm{d}n/\mathrm{d}x$
obtained from eqn.~(\ref{eqn:dijet1}) over that bin.
Thus \nbgvec{} is a function of the dijet background
parameters ${p_1, p_2, p_3}$. The number of expected signal events, \nsigvec, is evaluated based on MC simulation assuming the cross section of the model under test multiplied by the signal strength and including the effects of the systematic uncertainties described in section~\ref{sec:mcsig}.  The expected number of signal events is a function of  $\mu$ and the nuisance parameters modelling the systematic uncertainties on the signal. 

The compatibility of the data with the background-only expectation is quantified in terms of the local $p_0$, defined as the probability of the background-only model to produce an excess at least as large as the one observed and quantified with an ensemble of 500,000 background-only pseudo-experiments,  
while the global probability of an excess with a given local $p_0$ being the most significant excess to be observed anywhere in the search region is quantified with 100,000 background-only pseudo-experiments that take into account the mass ranges and overlapping event samples for the three channels.
The largest discrepancies, in the region around 2 \tev{} in figures~\ref{fig:djm_fit:wz}, \ref{fig:djm_fit:ww} and \ref{fig:djm_fit:zz}, lead to small $p_0$ values near that mass. The smallest local $p_0$ values in the $WZ$, $WW$, and  $ZZ$ channels  correspond to significances of $\locsig{}~\sigma$, $2.6~\sigma$, and $2.9~\sigma$  respectively.
Considering the entire mass range of the search (1.3--3.0 \tev) in each of the three search channels, the global significance of the discrepancy in the $WZ$ channel is \globsig{}~$\sigma$.

Exclusion limits at the 95\% confidence level are set following the $CL_s$ prescription~\cite{Read:2002hq}.

\subsection{Exclusion limits on new diboson resonances}

\begin{figure}[th!]
  \begin{center}
    \subfigure[]
              {
     \includegraphics[width=0.48\textwidth]{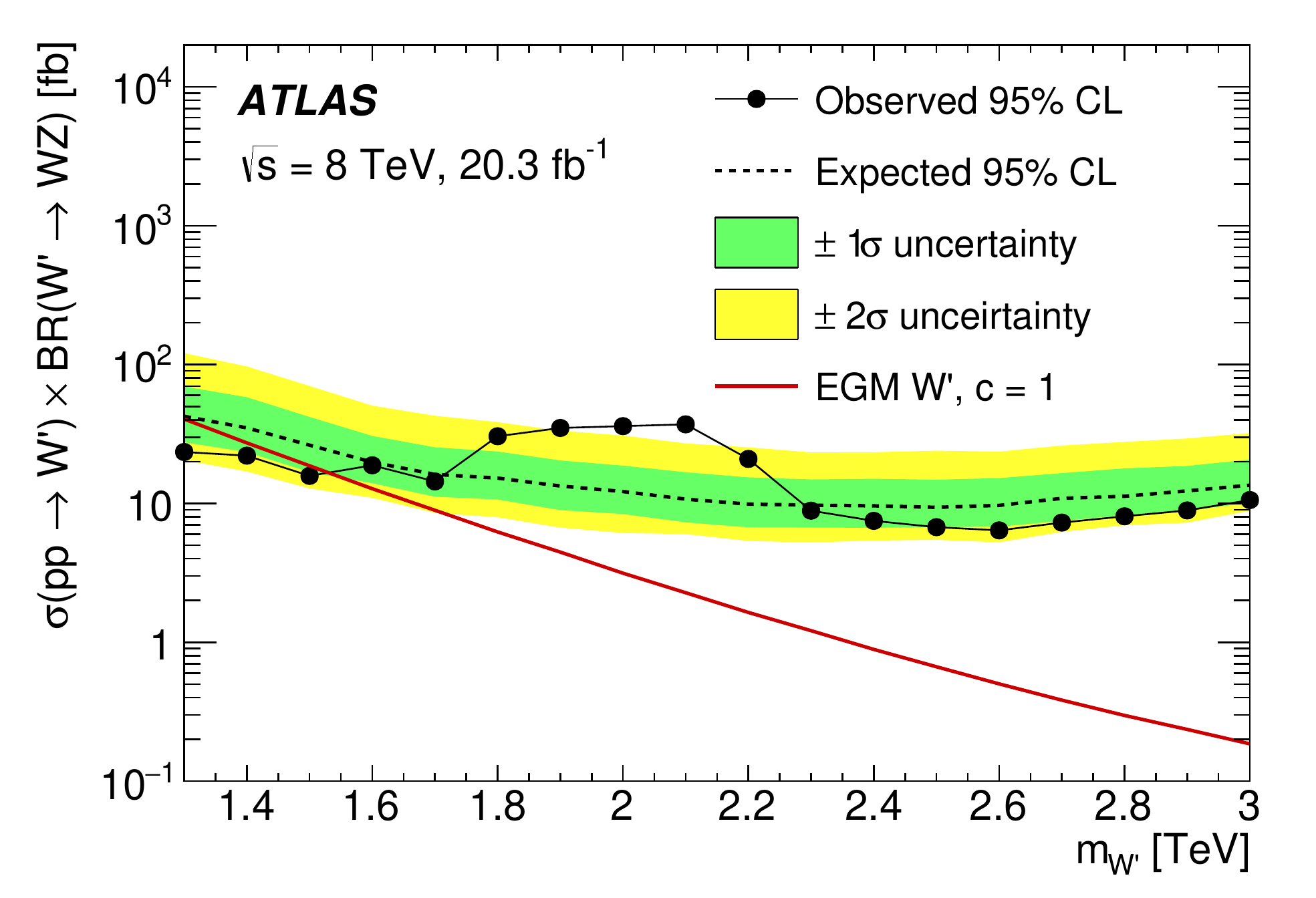}\label{fig:limit:WZ}
              }
    \subfigure[]
              {
     \includegraphics[width=0.48\textwidth]{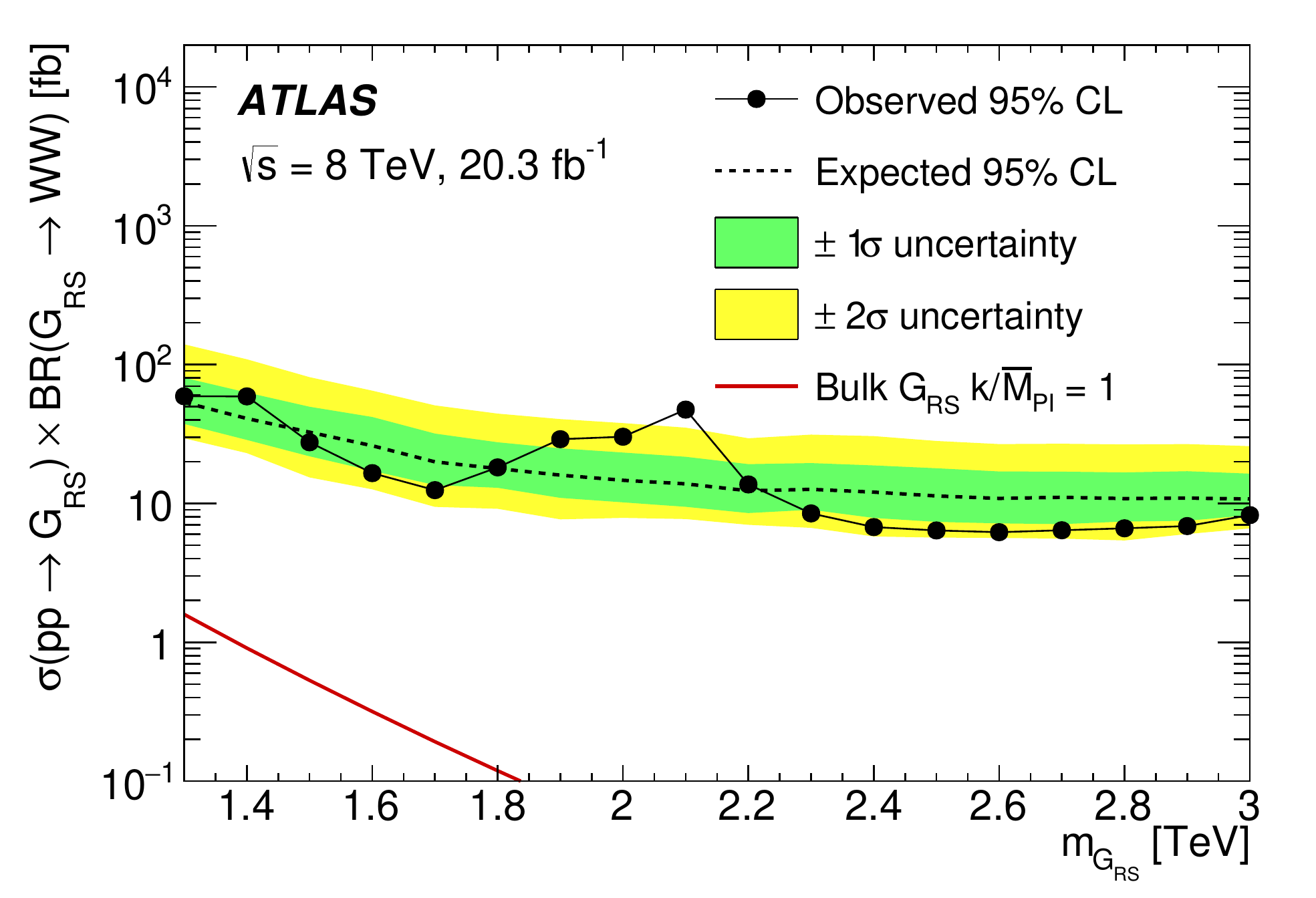}\label{fig:limit:WW}
              }
    \subfigure[]
              {
     \includegraphics[width=0.48\textwidth]{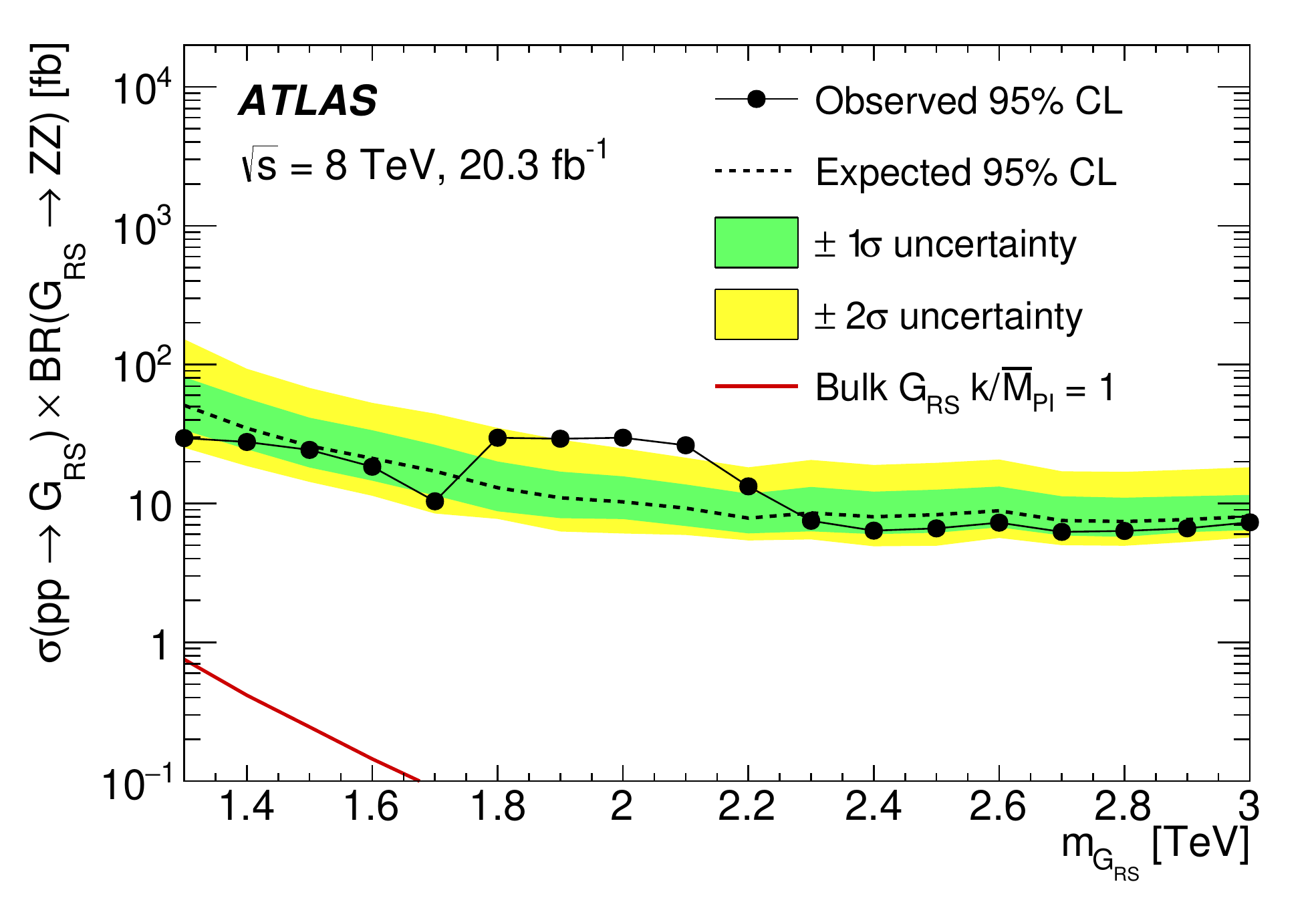}\label{fig:limit:ZZ}
              }
\end{center}
\vspace{-20 pt}
     \caption{Upper limits, at 95\% C.L., on the section times branching ratio limits for the $WZ$ window selection as a function of $m_{W'}$, and for the $WW$ window selection and the $ZZ$ window selections as a function of $m_{\grav}$. The solid red line in each figure displays the predicted cross section for the $W'$ or \grav{} model as a function of the resonance mass.}
\label{fig:limits}
\end{figure}

Limits on the production \crosssection{} times branching ratio of massive resonances are set in each diboson channel as a function of the resonance mass using the EGM $W'$ as a benchmark for the $WZ$ selection, and the bulk $G_{RS}$ model for the $WW$ and $ZZ$ selections. 
In most of the mass range, the observed limit is somewhat better than the expected limit, but in the region near 2 \tev{} the excess of events in the data leads to observed limits which are weaker than expected.  
Figure~\ref{fig:limit:WZ} shows the observed 95\% CL upper limits on the cross section times branching ratio on the EGM $W' \to WZ$ hypotheses as a function of the $W'$ mass. EGM $W' \to WZ$ for masses between 1.3 and 1.5 \tev{} are excluded at 95\% CL.  Figures~\ref{fig:limit:WW} and~\ref{fig:limit:ZZ} show the observed 95\% CL upper limits on the cross section times branching ratio for the bulk $G_{RS} \to WW$ and $ZZ$, respectively. The \crosssection{} times branching ratio for excited graviton production with the model parameters described in section~\ref{sec:sim} is too low to be excluded with the sensitivity of this measurement.

\section{Conclusions}
A search has been performed for massive particles decaying to $WW$, $WZ$, or $ZZ$ using  20.3 fb$^{-1}$ of  $\sqrt{s}=8 \tev$ $pp$ collision data collected at the LHC by ATLAS in 2012. This is the first ATLAS search for resonant diboson production in a fully hadronic final state and strongly relies on the suppression of the dijet background with a substructure-based jet grooming and boson tagging procedure.  The boson tagging selection includes different jet mass criteria to identify $W$ and $Z$ boson candidates and thus produces three overlapping sets of selected events for the searches in the $WW$, $WZ$, and $ZZ$ decay channels.
The most significant discrepancy with the background-only model occurs around 2 \tev{} in the $WZ$ channel with a local significance of \locsig{} $\sigma$ and a global significance, taking the entire mass range of the search in all three channels into account, of \globsig{}~$\sigma$.  

Upper limits on the production \crosssection{} times branching ratio of massive resonances are set in each diboson channel as a function of the resonance mass, using an EGM \wpwz\ as a benchmark for the $WZ$ channel, and an excited bulk graviton $G_{RS}$ to represent resonances decaying to $WW$ and $ZZ$.  A $W'$with EGM couplings and mass between 1.3 and 1.5 \tev{} is excluded at 95\% CL.  

\section*{Acknowledgements}


We thank CERN for the very successful operation of the LHC, as well as the
support staff from our institutions without whom ATLAS could not be
operated efficiently.

We acknowledge the support of ANPCyT, Argentina; YerPhI, Armenia; ARC, Australia; BMWFW and FWF, Austria; ANAS, Azerbaijan; SSTC, Belarus; CNPq and FAPESP, Brazil; NSERC, NRC and CFI, Canada; CERN; CONICYT, Chile; CAS, MOST and NSFC, China; COLCIENCIAS, Colombia; MSMT CR, MPO CR and VSC CR, Czech Republic; DNRF, DNSRC and Lundbeck Foundation, Denmark; IN2P3-CNRS, CEA-DSM/IRFU, France; GNSF, Georgia; BMBF, HGF, and MPG, Germany; GSRT, Greece; RGC, Hong Kong SAR, China; ISF, I-CORE and Benoziyo Center, Israel; INFN, Italy; MEXT and JSPS, Japan; CNRST, Morocco; FOM and NWO, Netherlands; RCN, Norway; MNiSW and NCN, Poland; FCT, Portugal; MNE/IFA, Romania; MES of Russia and NRC KI, Russian Federation; JINR; MESTD, Serbia; MSSR, Slovakia; ARRS and MIZ\v{S}, Slovenia; DST/NRF, South Africa; MINECO, Spain; SRC and Wallenberg Foundation, Sweden; SERI, SNSF and Cantons of Bern and Geneva, Switzerland; MOST, Taiwan; TAEK, Turkey; STFC, United Kingdom; DOE and NSF, United States of America. In addition, individual groups and members have received support from BCKDF, the Canada Council, CANARIE, CRC, Compute Canada, FQRNT, and the Ontario Innovation Trust, Canada; EPLANET, ERC, FP7, Horizon 2020 and Marie Skłodowska-Curie Actions, European Union; Investissements d'Avenir Labex and Idex, ANR, Region Auvergne and Fondation Partager le Savoir, France; DFG and AvH Foundation, Germany; Herakleitos, Thales and Aristeia programmes co-financed by EU-ESF and the Greek NSRF; BSF, GIF and Minerva, Israel; BRF, Norway; the Royal Society and Leverhulme Trust, United Kingdom.

The crucial computing support from all WLCG partners is acknowledged
gratefully, in particular from CERN and the ATLAS Tier-1 facilities at
TRIUMF (Canada), NDGF (Denmark, Norway, Sweden), CC-IN2P3 (France),
KIT/GridKA (Germany), INFN-CNAF (Italy), NL-T1 (Netherlands), PIC (Spain),
ASGC (Taiwan), RAL (UK) and BNL (USA) and in the Tier-2 facilities
worldwide.

\printbibliography
\newpage
\begin{flushleft}
{\Large The ATLAS Collaboration}

\bigskip

G.~Aad$^{\rm 85}$,
B.~Abbott$^{\rm 113}$,
J.~Abdallah$^{\rm 151}$,
O.~Abdinov$^{\rm 11}$,
R.~Aben$^{\rm 107}$,
M.~Abolins$^{\rm 90}$,
O.S.~AbouZeid$^{\rm 158}$,
H.~Abramowicz$^{\rm 153}$,
H.~Abreu$^{\rm 152}$,
R.~Abreu$^{\rm 30}$
% ...
\bigskip
\\
% ...
$^{11}$ Institute of Physics, Azerbaijan Academy of Sciences, Baku, Azerbaijan\\
% ...
$^{30}$ CERN, Geneva, Switzerland\\
% ...
$^{85}$ CPPM, Aix-Marseille Universit{\'e} and CNRS/IN2P3, Marseille, France\\
% ...
$^{90}$ Department of Physics and Astronomy, Michigan State University, East Lansing MI, United States of America\\
% ...
$^{107}$ Nikhef National Institute for Subatomic Physics and University of Amsterdam, Amsterdam, Netherlands\\
% ...
$^{113}$ Homer L. Dodge Department of Physics and Astronomy, University of Oklahoma, Norman OK, United States of America\\
% ...
$^{151}$ Institute of Physics, Academia Sinica, Taipei, Taiwan\\
$^{152}$ Department of Physics, Technion: Israel Institute of Technology, Haifa, Israel\\
$^{153}$ Raymond and Beverly Sackler School of Physics and Astronomy, Tel Aviv University, Tel Aviv, Israel\\
% ...
$^{158}$ Department of Physics, University of Toronto, Toronto ON, Canada\\
% ...
\end{flushleft}



\end{document}